\documentclass[10pt,letterpaper]{article}
\usepackage[top=1.0in,left=1.0in,right=1.0in,footskip=0.75in]{geometry}

\usepackage{amsmath,amssymb,amsthm,bbold}
\usepackage{changepage}
\usepackage{longtable}

\usepackage{times}

\usepackage{cite}
\usepackage{nameref}
\usepackage[bookmarks=false]{hyperref}
\PassOptionsToPackage{linktocpage}{hyperref}
\usepackage[table]{xcolor}
\usepackage{array}
\usepackage{graphicx}

\newcommand{\blank}[1]{}
\newcommand{\ha}{\hat{a}}
\newcommand{\adag}{\hat{a}^\dagger}
\DeclareMathOperator{\logit}{logit}

\date{}

\begin{document}

\begin{flushleft}
{\Large
\textbf\newline{Widespread bursts of diversification in microbial phylogenies} 
}
\newline
\\
Alice Doucet Beaupr\'e\textsuperscript{1,2*$\ddagger$},
James P. O'Dwyer\textsuperscript{2,3$\ddagger$},
\\
\bigskip
\textbf{1} Program in Ecology, Evolution, and Conservation Biology, University of Illinois, Urbana, Illinois, USA
\\
\textbf{2} Department of Plant Biology, University of Illinois, Urbana, Illinois, USA
\\
\textbf{3} Carl R. Woese Institute for Genomic Biology, University of Illinois, Urbana, Illinois, USA
\\
\bigskip

$\ddagger$ These authors contributed equally to this work.

* doucetb2@illinois.edu

\end{flushleft}
\section*{Abstract}
Until recently, much of the microbial world was hidden from view. A global research effort has changed this, unveiling and quantifying microbial diversity across enormous range of critically-important contexts, from the human microbiome, to plant-soil interactions, to marine life. Yet what has remained largely hidden is the interplay of ecological and evolutionary processes that led to the diversity we observe in the present day. We introduce a theoretical framework to quantify the effect of ecological innovations in microbial evolutionary history, using a new, coarse-grained approach that is robust to the incompleteness and ambiguities in microbial community data. Applying this methodology, we identify a balance of gradual, ongoing diversification and rapid bursts across a vast range of microbial habitats. Moreover, we find universal quantitative similarities in the tempo of diversification, independent of habitat type.

\section*{Introduction}

Large scale microbiome sampling and  sequencing~\cite{Gilbert2014,hmp,amaral2010global,zinger_global_2011} have documented global microbial diversity with unprecedented scope and resolution. The tools currently applied to these data allow us to quantify the amount and type of diversity found in microbial communities~\cite{meyer_metagenomics_2008,schloss_introducing_2009,caporaso_qiime_2010}, and yet we know remarkably little about the the underlying community dynamics and tempo of diversification that generated the biodiversity we observe. This gap in our knowledge calls out for robust new ecological and evolutionary theories that will allow us to connect mechanisms to observed patterns~\cite{Costello2012}.

\smallskip

To address this challenge, we introduce a new methodology to bridge the gap between biological process and observed microbial biodiversity. Our approach leverages the inference of dynamical processes from evolutionary trees, previously applied to understand large-scale evolutionary structure\cite{Morlon2011fossil,may_bayesian_2016,hohna_tess_2016, maddison_estimating_2007,fitzjohn_estimating_2009,etienne_diversity-dependence_2012, rabosky_automatic_2014}, and also the dynamics of viral populations on shorter timescales\cite{grenfell2004unifying,Neher2013,kuhnert_simultaneous_2014,kuhnert_phylodynamics_2016}. We also incorporate the recent identification of bursts of diversification in microbial phylogenies~\cite{odwyer2015phylo}. The result is a model which includes traditional, slow processes for gradual speciation (one lineage goes to two lineages, which we call the `birth' of a lineage) and extinctions (one lineage disappears, which we call `death'), together with a third set of mechanisms, incorporating the process of ecological innovation potentially followed by radiative diversification.

\smallskip

We apply our framework to data spanning 13,500 individual samples, 56 habitat types, and 29 biomes~\cite{Gilbert2014}, finding a previously unidentified balance of fast and slow evolutionary processes in these data, and a tendency towards universal behaviour in the quantitative description of diversification.  We cannot directly quantify the traits and their changes through time that may have led to a given combination of rapid and gradual processes, but our results are strongly suggestive of a centre-ground in the long-standing debate over phyletic gradualism versus punctuated equilibrium~\cite{gould1977punctuated}.

\smallskip

\section*{Materials and methods}
\subsection{Coarse-grained phylodynamics}

Our knowledge of the diversification of a group of organisms is often characterised by the branching of their evolutionary lineages, reconstructed using genetic sequence data sampled in the present day.  We can think of an evolutionary tree, also known as a phylogeny, as the input data for inferring theoretical models of diversification, a methodology known as phylodynamics~\cite{grenfell2004unifying,maddison_estimating_2007,fitzjohn_estimating_2009,Morlon2011fossil,etienne_diversity-dependence_2012,rabosky_automatic_2014,kuhnert_simultaneous_2014,kuhnert_phylodynamics_2016,may_bayesian_2016,hohna_tess_2016}. This approach is usually thought of in terms of speciation and extinction rates. But in recent work~\cite{odwyer2015phylo} we identified patterns of bursts in the branching in microbial phylogenies. Therefore, in addition to the traditional speciation-extinction process, we also parametrize fast (but brief) bursts of diversification. In these burst processes, we propose that a lineage will undergo a much faster rate of diversification, $\sigma$, but for a very short time, $\tau$. The origin of a burst could e.g. be a key functional change that opens the opportunity for a rapid radiation~\cite{schluter2000ecology,rainey1998adaptive}, or a disturbance which opens up a new habitat to be colonized. In the following we will refer interchangeably to this third process as bursts or innovations.

\begin{figure}
\centering
\includegraphics[scale=1.2]{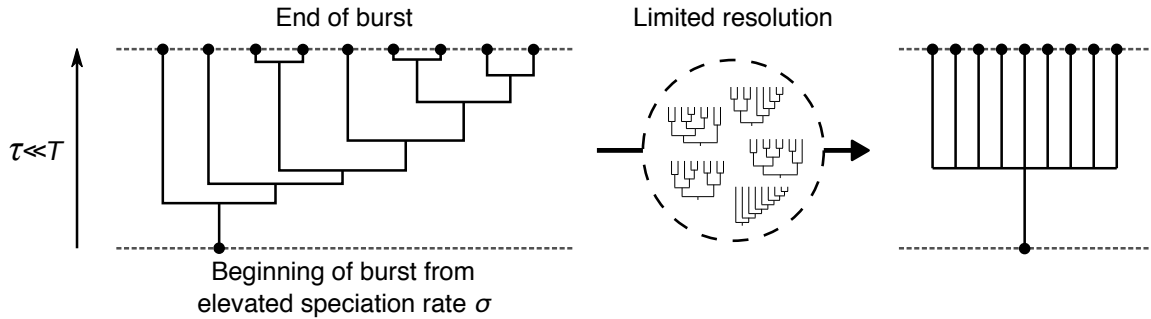}
\caption{\textbf{Uncertainty generates polytomies.} Periods of fast diversification leave little to no signal in sequences with a limited number of base pairs, meaning that we cannot always distinguish between different possible orderings of diversification events using short sequences. These ambiguities often rightfully show up in bootstrap consensus trees or during calibration. Considering the tree as an ensemble of transitions happening over a coarse-grained time interval alleviates this issue and allows the inference of effective parameters associated with faster processes.}
\label{fig:burst}
\end{figure}

\begin{figure}
\centering
\includegraphics[scale=0.85]{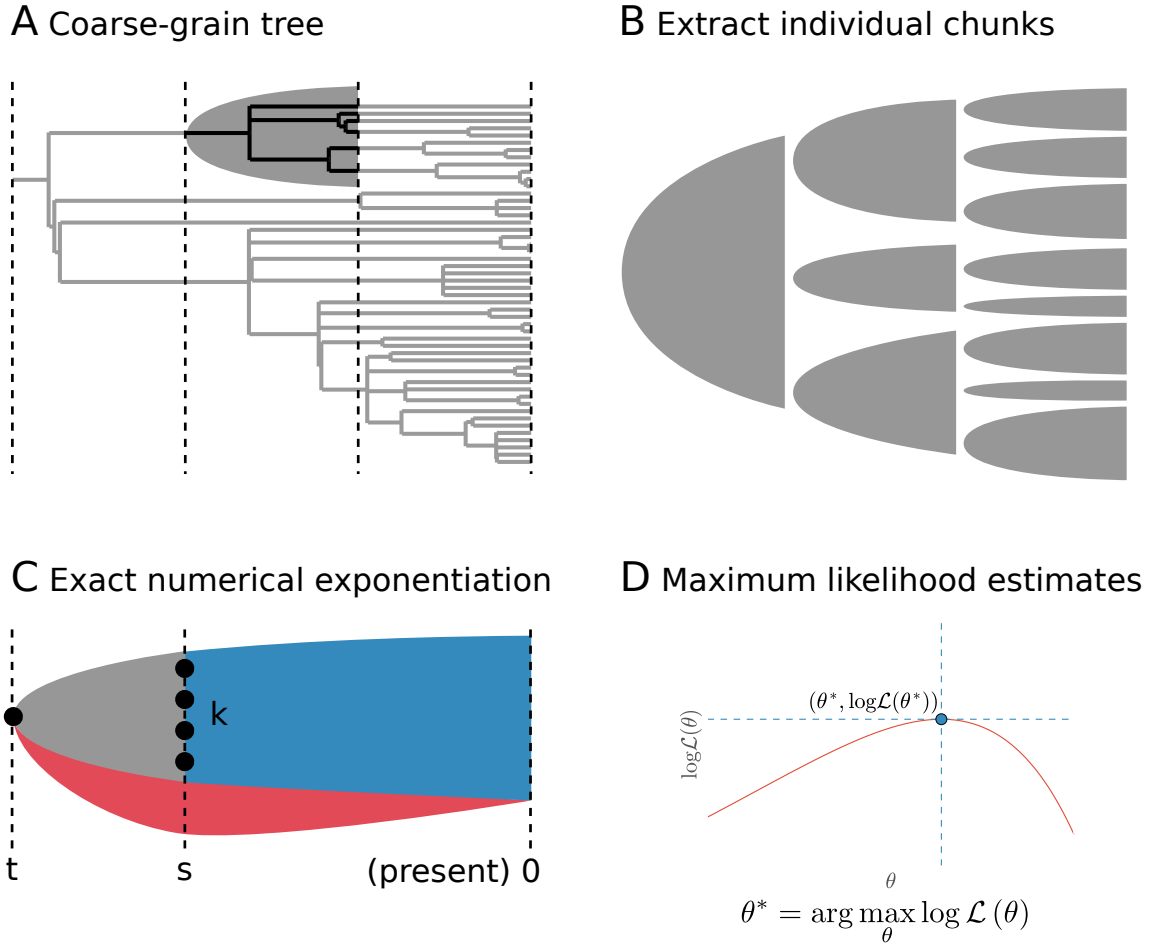}
\caption{\textbf{Coarse-grained phylodynamics inference pipeline.} ({\bf A}) We cut a phylogeny of depth $T$ at a predetermined number of slices $K$, based on sequence length. Slice boundaries isolate pieces of the tree that begin with a single lineage further in the past and end at $k\geq 1$ lineages at the next slice boundary closer to the present. The isolated dark-grey piece above begins in the past at $2T/3$ with one lineage and ends closer to the present at $T/3$ with $k=6$ lineages. ({\bf B}) Once sliced, the tree decomposes into many isolated pieces, or chunks.  Each chunk is identified by $(t, s, k)$: its origin at $t$, its end at the next slice boundary $s$, and the number of observed lineages $k$ at time $s$. ({\bf C}) To each chunk corresponds a conditional transition probability which traces over all unobserved, extinct lineages (red), and over all histories consistent with observed lineages (gray), all of them with extant descendants (blue). ({\bf D}) We form the full log-likelihood of observing a given tree by adding together all individual chunk log-likelihoods. Parameter estimates (denoted generically by $\theta^*$) are obtained by maximizing the log-likelihood.}
\label{fig:pipeline}
\end{figure}

There are two problems with inferring the parameters of this generalized innovation process when $\sigma$ is large. First, there will be parts of any reconstructed phylogeny where we may not have enough information in our sequence data to distinguish the true ordering of very fast branching events, as shown in Figure~\ref{fig:burst}.  Even if we did have longer sequences, there is always a speed limit on what kinds of process we can accurately infer from these data, making these rates hard to infer. Second, in any realistic evolutionary history we would expect many different rates $\sigma$, corresponding to the idiosyncrasies of individual events.  In this new approach, we (partially) bypass both difficulties by applying a method of coarse-graining, where we decompose a sample phylogeny of age $T$ into $K$ slices of width $T/K$ (Fig.~\ref{fig:pipeline} Panel A).  In a coarse-grained phylogeny we are no longer trying to resolve down to each binary split in the tree---what we have access to are the `chunks' of diversification between time slices (Fig.~\ref{fig:pipeline} Panel B), which define equivalence classes of binary trees.

Surprisingly, there is a way to bypass this speed limit, by leveraging the distribution of sizes of these (apparent) bursts of branching. Even though we can't resolve phylogenies down to the shortest timescales, this distribution still carries information about the parameters of the innovation process. The catch is that we cannot distinguish between different values of $\sigma$ and $\tau$ independently, as the distribution of burst sizes is a geometric distribution which only depends on the product of diversification rate and diversification time, $\sigma\tau$.  By looking at the evolutionary history through a blurred lens, we therefore collapse a multi-parameter family of models into a single parameter, reminiscent of the loss of information under coarse-graining in physics---so that at a sufficiently coarse resolution, many different fine-scale models map onto the same effective theory.

The final steps in our pipeline (Fig.~\ref{fig:pipeline} Panels C and D) involve computing likelihoods for a given model of innovation using the chunks defined by coarse-graining. We considered three distinct models in this preliminary work:  the traditional speciation-extinction process (abbreviated BD, for birth and death); our basic speciation, extinction and individual burst process (BDI), which assumes a single value of the product $\sigma\tau$; and finally what we call speciation, extinction and heterogeneous bursts (BDH), where we compound the innovation processes using a distribution over the product $\sigma\tau$. BDI and BDH offer an agnostic and parsimonious alternative to models with time-varying~\cite{Morlon2011fossil,may_bayesian_2016,hohna_tess_2016}, trait-dependent~\cite{maddison_estimating_2007,fitzjohn_estimating_2009}, and diversity-dependent~\cite{etienne_diversity-dependence_2012, rabosky_automatic_2014} models. 

The likelihoods are then calculated via numerical solutions of a master equation, described below, for the probability that an initial lineage (at the beginning of any chunk) will have branched into $k\ge 1$ lineages at the end of the chunk conditional on each lineage having at least one extant descendant. Each chunk represents a properly weighted sum over many histories compatible with uncertain tree structures and therefore helps us bypass the need for an accurate estimate of branch lengths. 

Microbial phylogenies usually span enormous amounts of evolutionary time with relatively short sequences and may suffer from a lack of phylogenetic signal sufficient to reconstruct early ancestral states, which in turn imposes an horizon deep in the tree beyond which topological inaccuracies become inevitable~\cite{mossel_how_2005}. Furthermore, slices closer to the present contribute much more weight in the total likelihood than those from an otherwise noisy and uncertain past because they contain more chunks. For example, in Fig.~\ref{fig:cumulchunkdbn} below, consider the intersection of the cumulative distribution and the y-axis equals the number of chunks in that slice. We see that the slice closest to the present (top-left panel) contributes at least an orders of magnitude more chunks than the earliest ones (panels in bottom two rows) where low phylogenetic signal potentially degrades the quality of the input phylogeny.

What happens if we apply this methodology to a perfect, fully-resolved tree? Reassuringly, we show in the Supplementary Information that in this limit and using a very large number of slices our inference using the BD model exactly recapitulates previous approaches~\cite{thompson_human_1975,yang_bayesian_1997,gernhard_conditioned_2008,stadler_incomplete_2009,Morlon2011fossil}. On the other hand, in cases where we do have limited resolution due to short sequences, our method extends the current applicability of BD models by allowing the inference of rate parameters over incompletely-resolved phylogenies. We can then distinguish between our three nested models using likelihood ratio testing, and we also perform an exact goodness-of-fit test by comparing a given empirical tree to an ensemble of typical trees generated by the model and its parameter estimates constrained to the same empirical size and depth (see Supplementary Information and Fig.~S4).

\subsection{Data availability}
All amplicon sequence data and metadata have been made public through the data portal (qiita.microbio.me/emp) and all accession numbers to the European Nucleotide Archive (www.ebi.ac.uk/ena) can be found in Table~S1. All processed individual phylogenetic trees used in this study will be made public prior to publication.
\subsection{Code availability}
Custom code for the tree coarse-graining, maximum likelihood inference, and Markov Chain Monte Carlo (MCMC) goodness-of-fit test is stored in a private GitHub repository and will be made public prior to publication.

\subsection{Dataset and data preparation}
As an initial application, we considered the Earth Microbiome Project (EMP) 10K (as of 2015), drawing from 5.59 million representative 16S rRNA sequences (open reference 97\% similarity cut-off, V4 region\cite{rideout_subsampled_2014}) gathered from 56 different studies and breaking into 13,500 separate samples~\cite{Gilbert2014}. Table~S1 lists these studies, together with a short description as it appears in the EMP metadata, and a clickable URL to their Qiita entries and accession number or study website. These samples span 29 biomes and 56 habitats, including host-associated habitats from 93 different types of host. The EMP dataset is available on the project website (\href{http://www.earthmicrobiome.org}{www.earthmicrobiome.org}) as single large phylogenetic tree reconstructed using FastTree~\cite{Price2010,price2010fasttree} and representative sequences for each taxonomic unit from all habitats through the Qiita data portal (\href{https://qiita.ucsd.edu}{qiita.ucsd.edu}).

The first critical question to address is to what extent we should aggregate samples when applying our phylodynamic inference. Should all human-associated samples be aggregated, or all soil samples? We know that the ancestors of present-day organisms in each sample were unlikely to be co-located, experiencing the same environmental context for their entire evolutionary history.  On the other hand, we don't want simply to pool all samples together; microbial life from sufficiently different habitats is likely to have different underlying evolutionary and ecological processes. For this study we took the approach of inferring parameters sample by sample, but it would be straightforward to aggregate any particular group of sampled data. For example, questions pertaining to phylogeography may necessitate the use of a spatially hierarchical aggregation scheme. We began with the large EMP phylogeny containing all representative sequences and we transformed it into a chronogram using the method of mean path lengths~(MPL)~\cite{britton_phylogenetic_2002,britton_estimating_2007} weighted by relative operational taxonomic unit abundances. We chose this method because likelihood and semi-likelihood approaches for ultrametrizing phylogenies do not scale well for trees with more than 1,000 to 10,000 branches or leaves. MPL on the other hand requires only two traversals of the tree and therefore scale linearly with the size of the tree---ultrametrizing the full EMP phylogeny with MPL takes about 5 minutes on a consumer laptop. Alternatively supertree methods\cite{marin_timetree_2016} could also be used for transforming in a reasonable amount of time very large phylogenies into a chronogram.

This second step is necessary because, like all phylodynamic approaches, our models are based on changes through time, whereas our reconstructed trees are based on sequence divergence. Transforming the tree into a chronogram puts the two on the same footing. We did not need to calibrate the root of the full EMP tree in actual Myr or Byr because only the dimensionless rates and exponents matter for the quantitative comparison and characterization of gradual (slow) vs bursty (fast) evolution. From this large phylogeny we then pruned a smaller subtree for each individual sample. These individual sample phylogenies form the input data for our analyses.
\subsection{Model definitions}
All models considered in this study (BD, BDI, and BDH) are constructed from four building blocks which create and destroy lineages which we will represent by the letter $L$, namely the stochastic birth process
\begin{equation}\label{eq:birtharrow}
\text{Birth (B):}\quad L \overset{b}{\longrightarrow} 2L,
\end{equation}
with per-lineage birth/speciation rate $b$, the death process
\begin{equation}\label{eq:deatharrow}
\text{Death (D):}\quad L \overset{d}{\longrightarrow} \emptyset,
\end{equation}
with per-lineages death/extinction $d$, the innovation process
\begin{equation}\label{eq:innovarrow}
\text{Innovation (I):}\quad L \overset{\rho g_x(k)}{\longrightarrow} kL,
\end{equation}
where $\rho$ is the per-lineage innovation initiation rate $\rho$ and transition probabilities $g_x(k) = (1-x)x^{k-1}$ for $k \ge 1$ and 0 otherwise. The parameter $x = 1 - e^{-\sigma\tau}$ characterizes the geometric burst-size distribution. Finally we compound the innovation process with a beta distribution over $x$. The beta distribution is parametrized by its precision $s$ and mean $m$ to obtain the heterogeneous innovation process 
\begin{equation}\label{eq:hinnovarrow}
\text{Heterogeneous Innovation (H):}\quad L \overset{\eta h_{s,m}(k)}{\longrightarrow} kL,
\end{equation}
where $\eta$ is the per-lineage heterogeneous innovation initiation rate and transition probabilities $h_{s,m}(k) = s(1-m)\frac{\Gamma(s)\Gamma(sm+k-1)}{\Gamma(sm)\Gamma(s+k)}$ for $k\ge 1$ and 0 otherwise. The beta distribution is sometimes parametrized by two shape parameters, $\alpha$ and $\beta$, in our case controlling the behavior of the parameter $x$, now a random variable, around 0 and around 1, respectively. In terms of those shape parameters the precision $s = \alpha + \beta$ and mean $m = \alpha/(\alpha + \beta)$.

For each process there is an associated stochastic generator which encodes the instantaneous transition rates between different number of lineages presented in equations (\ref{eq:birtharrow}) through (\ref{eq:hinnovarrow}). Generators for the above processes are respectively given by
\begin{equation}
\text{(B):}\quad \mathcal L_B(b) = b\left( z - 1\right)z\partial_z,
\end{equation}
\begin{equation}
\text{(D):}\quad\mathcal L_D(d) = d\left(1 - z\right)\partial_z,
\end{equation}
\begin{equation}
\text{(I):}\quad\mathcal L_I(\rho, x) = \rho x \frac{\left(z - 1\right)z}{1- x z}\partial_z,\text{ and}
\end{equation}
\begin{equation}
\text{(H):}\quad\mathcal L_H(\eta, s, m) =\eta m~{}_2F_1\left(1, sm+1, s + 1; z\right)\left(z - 1\right)z\partial_z.
\end{equation}
To combine models one adds generators together. For example the generator of the BDH model is written $\mathcal L_{BDH} = \mathcal L_B + \mathcal L_D + \mathcal L_H$ with associated parameter set $\theta_{BDH} = \left\lbrace b, d, \eta, s, m\right\rbrace$.
\subsection{Coarse-grained phylogeny likelihood}
The coarse-graining/slicing operator $\mathcal R_K$ decomposes a sample phylogeny $\mathcal{T}$ of age $T$ into slices of width $T/K$. The resulting coarse-grained phylogeny is characterized by a multiset of chunks $\mathcal{T}_K = \mathcal R_K\left[\mathcal{T}\right] = \lbrace (t_i,s_i,k_i)\rbrace_{i\in\mathcal I[\mathcal{T}_k]}$, with $\mathcal I$ the index map given some arbitrary tree traversal. The log-likelihood of observing $\mathcal{T}_K$ under a given models with parameter set $\theta$ is written
\begin{equation}
\log \text{Pr}\left[\mathcal{T}_K\vert \theta\right] = \sum_{i\in\mathcal I[\mathcal{T}_k]} \log \phi^{(k_i)}(t_i, s_i\vert \theta)
\end{equation}
and maximum likelihood parameter estimates
\begin{equation}
\theta^* = \arg\max_\theta \log \text{Pr}\left[\mathcal{T}_K\vert \theta\right].
\end{equation}
\subsection{Chunk likelihoods}
The expression for individual chunk likelihoods is given by
\begin{equation}	
\phi^{(k)}(t, s\vert \theta) = \frac{1}{k!}\left.\partial^k_y\Phi_{t,s}(y\vert \theta)\right\vert_{y=0}.
\end{equation}
The chunk generating function
\begin{equation}
\Phi_{t,s}(y\vert \theta) = \frac{\mathcal U_{t-s}\left(\mathcal U_s(0\vert\theta) + y\left(1 - \mathcal U_s(0\vert\theta)\right)\vert\theta\right) - \mathcal U_t(0\vert\theta)}{1 - \mathcal U_t(0\vert\theta)},
\end{equation}
where $\mathcal U_\tau(z\vert\theta)$ is the probability generating function
\begin{equation}
\mathcal U_\tau(z\vert\theta) = \sum_{k=0}^\infty P\left(k\vert \tau, \theta\right)z^k,
\end{equation}
solution to the master equation
\begin{equation}
\partial_\tau \mathcal U_\tau(z\vert\theta) = \mathcal L(\theta)~\mathcal U_\tau(z\vert\theta)
\end{equation}
with initial condition $U_0(z\vert\theta) = z$. The generator $\mathcal L(\theta)$ is one of $\mathcal L_{BD}(b, d)$, $\mathcal L_{BDI}(b, d, \rho, x)$, or $\mathcal L_{BDH}(b, d, \eta, s, m)$. The subtraction of $\mathcal U_t(0\vert\theta)$ in the numerator and normalization by $1-\mathcal U_t(0\vert\theta)$ in the denominator account for conditioning on non-extinction of at least one observed extant lineage. Details on the derivation, generalization to incomplete sampling of lineages, and numerical computation of the above quantities can be found in the Supplementary Information.
\subsection{Model comparison and goodness of fit}
The BD model is nested into BDI, and BDI into BDH, therefore we can perform a likelihood ratio test by comparing the $D$-statistic
\begin{equation}
D =  2\left(\log \text{Pr}\left[\mathcal{T}_K\vert \theta^*_{alt}\right] - \log \text{Pr}\left[\mathcal{T}_K\vert \theta^*_0\right]\right)
\end{equation}
against a $\chi^2_{\text{ddof}}$ distribution with number of degrees of freedom $\text{ddof} = \vert \theta^*_{alt}\vert - \vert \theta^*_0\vert$. For $(\mathcal{H}_{alt}, \mathcal{H}_0) = (\text{BDI}, \text{BD})$ we have $\text{ddof}= 2$, and for $(\mathcal{H}_{alt}, \mathcal{H}_0) = (\text{BDH}, \text{BDI})$ we have $\text{ddof} = 1$.

We perform an exact goodness of fit test by sampling the $G$-statistic over the space of possible input coarse-grained phylogenies constrained by the depth and number of leaves of the empirical sample phylogeny. This statistic is given by
\begin{equation}
G(\mathcal{T}_K) = 2 \sum_{\sigma=0}^{K-1}\sum_{k\in\mathcal{K}_\sigma} n^\sigma_k \log\frac{n^\sigma_k}{N_\sigma \phi^{(k)}(t_\sigma, s_\sigma)},
\end{equation}
where the first sum runs over slices and the second one over all chunk sizes found in slice $\sigma$. In the summand $n^\sigma_k$ represent the number of chunks of size $k$, and $N_\sigma = \sum_{k} n^\sigma_k$ the total number chunks, in slice $\sigma$. The $G$-statistic corresponds to the information divergence between the empirical and the theoretical chunk size distribution. The exact goodness of fit statistic is given by the fraction of coarse-grained phylogeny with $G(\mathcal{T}'_K)$ greater than the empirical value $G(\mathcal{T}_K)$ where $\mathcal{T}'_K$'s are sampled using the MCMC Metropolis-Hasting algorithm poised at the maximum likelihood parameter estimates. We describe the algorithm and coarse-grained proposal distribution in the Supplementary Information. This step is necessary because the number of degrees of freedom of trees with fixed depth, fixed number of leaves, and fixed number of slices is unknown.

\section*{Results}
\begin{figure*}
\centering
\includegraphics[scale=0.26]{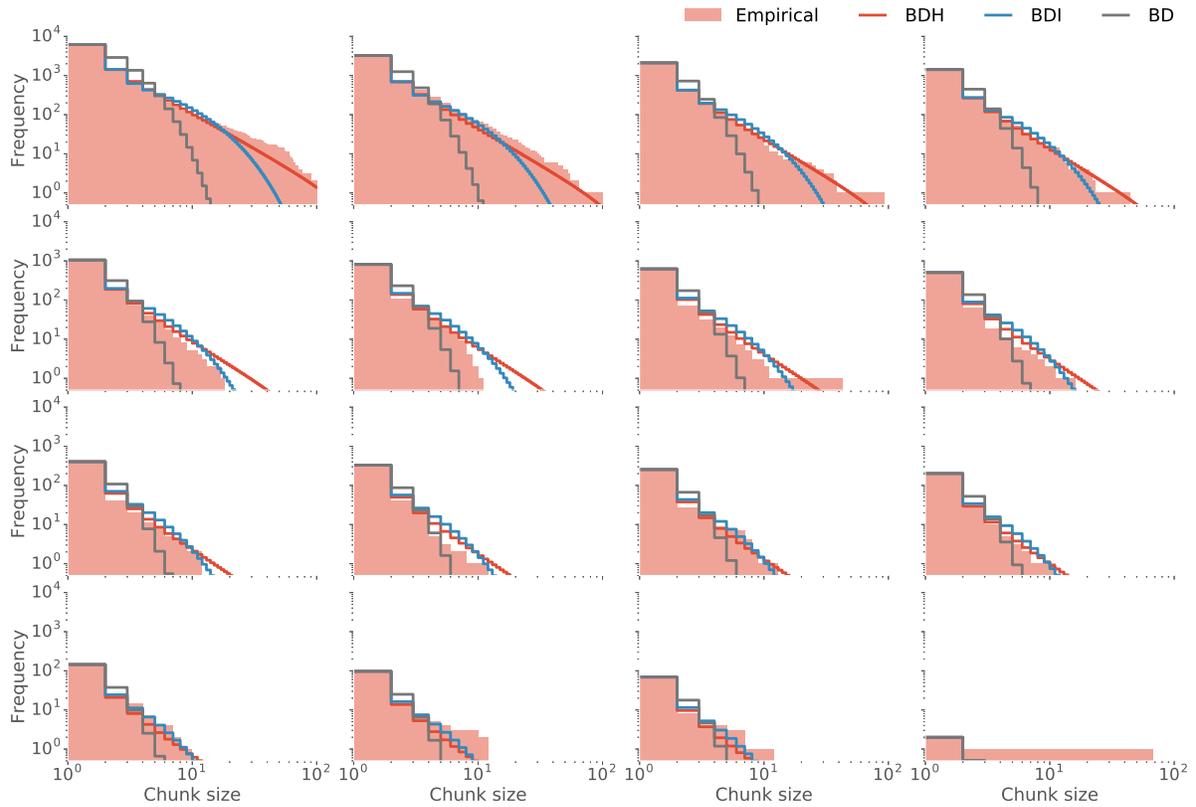}
\caption{\textbf{Cumulative chunk distributions across slices are approximately distributed as a power law.} The empirical tree for this particular case comes from the gut microbiota of captive colobine primates\cite{Amato2016225} (stomach mucosa of a captive Northern Douc, sample ID 45300SDZ3.F1.Pnem.stom.609719, study MetSan, see Table S1) The upper left plot stands for the distribution within the slice closest to the present, i.e. including the leaves. Time increases toward to past from left to right and from top to bottom. The red line represents the maximum likelihood distribution fitted using the heterogeneous innovation model, the blue line the basic innovation model, and the gray line the standard speciation-extinction model.  Note that all time slices use the same  parameter estimates for a given models---i.e. the model is fitted using the whole tree, not tuned slice-by-slice.}
\label{fig:cumulchunkdbn}
\end{figure*}

We rejected the BD hypothesis in favour of innovation (the BDI or BDH model) according to a likelihood ratio test for nested models (BD~$\subset$~BDI~$\subset$~BDH) at significance level $p~=~2\times 10^{-11}$, corresponding to a Bonferroni corrected $5\sigma$ family-wise error rate level $\alpha~=~2.7\times10^{-7}$. In cases where BD was rejected, we use the maximum likelihood estimates for the parameters of the BDI or BDH process to analyze the balance of slow and fast processes (birth rates vs innovation rates) and the phylogenetic signature of fast processes themselves (the distribution of burst sizes). Figure~S1 shows that BD was rejected in favour of BDI in 98\% of samples. Subsequently, our basic innovation model was rejected in favour of heterogeneous innovation in 80\% of samples. To understand why BDH is clearly selected in many cases it is instructive to look at a typical example. Figure~\ref{fig:cumulchunkdbn} shows how BDH captures the fatter tail of the empirical chunk size distribution across all slices of a coarse-grained phylogeny, while BD (and to an extent BDI) fail. Nonetheless, Fig.~S4 shows that for 57\% of samples BDI and BDH are sufficient to recapitulate the phylogeny, while for 43\% of samples both BDI and BDH fail the goodness-of-fit test, which suggests that we need a more complex model of innovation to account for them.

\begin{figure*}
\centering
\includegraphics[scale=0.45]{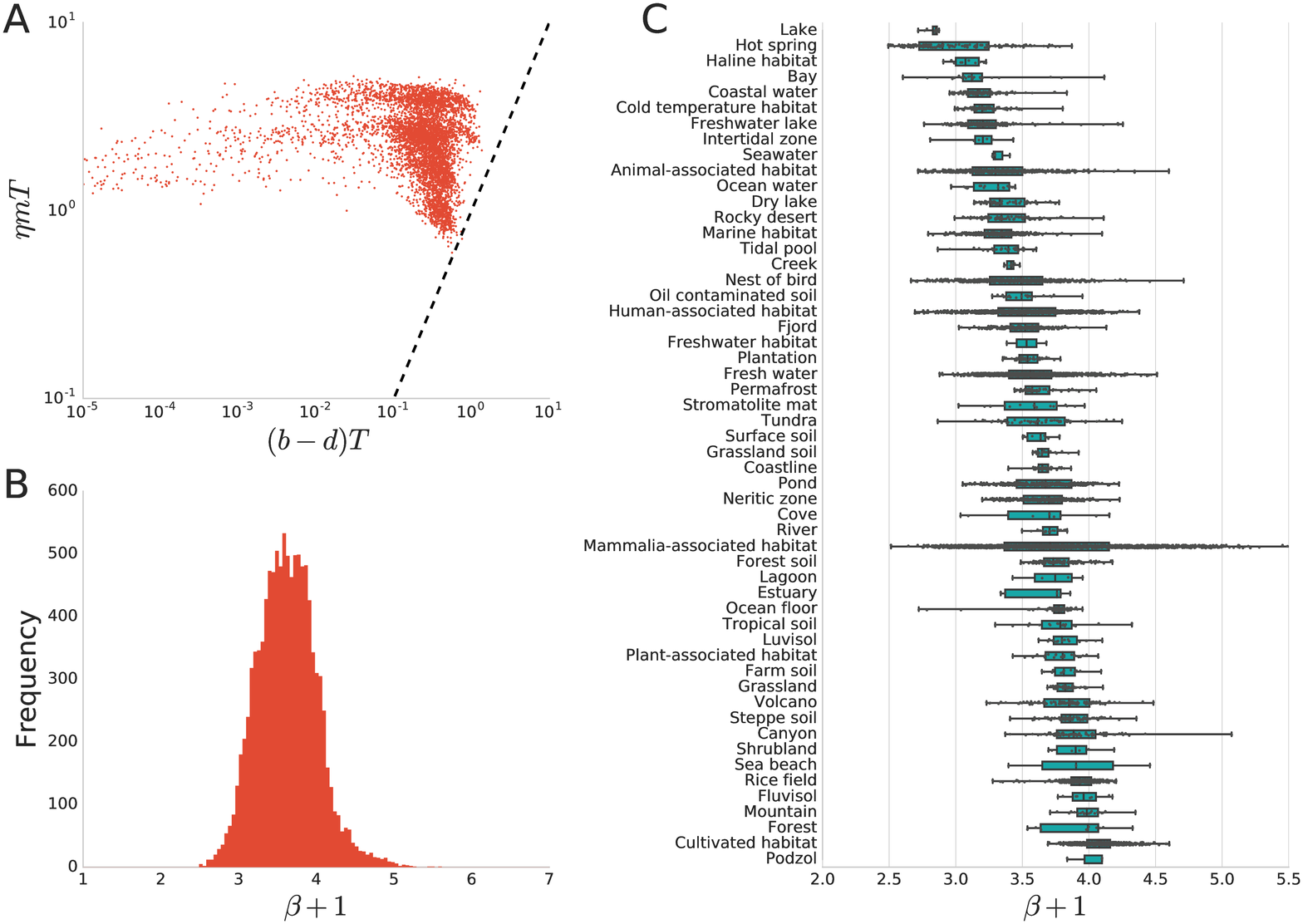}
\caption{\textbf{Distribution of  parameter estimates for the heterogeneous innovation model.} ({\bf A}) Each dot represents the effective rate of initiation of fast bursts vs the net rate of gradual diversification of an individual sample. We obtain comparable dimensionless effective rates by multiplication with the age of the root $T$ in each sample. ({\bf B}) The histogram shows the distribution of the BDH exponent $\beta+1$ characterizing the power law tail of heterogeneous innovation burst sizes. Smaller inferred values of $\beta+1$ imply that the burst size distribution has a heavier tail, and we identify a clustering of values across habitats.  ({\bf C}) Each dot represent the exponent $\beta+1$ of individual samples broken across different types of biome. Biomes are sorted according the their median. Boxes and whiskers indicate quartile limits.}
\label{fig:fastslow}
\end{figure*}

Beyond looking at single studies, our results can be expressed in terms of three important messages.  First, the mixture of bursty (in the form of innovation events) and gradual diversification is clearly preferred over pure gradualism in the vast majority of our samples. To explain the dynamics implied by reconstructed evolutionary trees, we need both ongoing, slow diversification, and bursts of faster diversification that last for a relatively short time. Second, even though fast processes by definition produce more lineages per unit time while they are in play, the initiation of bursts of any size is also more common than slow gradualism in samples that show evidence of heterogeneous innovation. Finally, in the right hand panel of Fig.~\ref{fig:fastslow} and in Fig.~S2-S3, we document the distribution of parameters controlling the shape of the burst size distribution for the heterogeneous innovation process. The distribution is beta-geometric and at large burst size $k$ this distribution behaves as $\sim k^{-(\beta+1)}$ where $\beta = s(1 - m)$. The effective exponent $\beta+1$ of this power law is clustered with median $3.63$ and quartile coefficient of dispersion $(Q_3-Q_1)/(Q_3+Q_1)=0.07$, a surprisingly narrow range of values, independent of the other estimated parameter values, the habitat, or the study. The apparent universality has echoes in recent work~\cite{Neher2013,odwyer2015phylo}, and in the long history of studying scaling in evolutionary history, for example in the number of species per genus in a given taxonomic group~\cite{rosenzweig_species_1995}. Our current analysis goes beyond documenting these patterns, by connecting this universality to a mechanistic interpretation.

\section*{Discussion}
We have introduced a new methodology to interpret what diversity in environmental sequence data can tell us about the ecological and evolutionary processes that shaped it. The key theoretical step in our new method is to recognize that faster diversification processes which appear intermittently and last only for a short time, still leave a signature in imperfectly reconstructed phylogenies. This signature persists even when the quality and length of our sequence data and consequent resolution of the phylogeny is relatively low compared to the timescale of the processes. Combining this realization with existing methods for inferring slower gradual processes from phylogenies, we were able to quantify the balance of fast and slow processes, and the parameter values that best describe the structure and distribution of burst sizes. Our conclusions in applying this to a large dataset encompassing heterogeneous habitats are stark: we almost always need these heterogeneous faster processes to complement gradual diversification in order to explain these data, and the parameters that best explain observed structure of evolutionary trees are surprisingly universal across studies and environmental context. 

\section*{Conclusion}
These results raise many questions, and open a number of doors for future investigation. Perhaps the primary open question is: what biological changes cause the bursts we observe in empirical trees?  Are these genuinely due to innovations, where an adaptation opens the door to many further adaptations~\cite{schluter2000ecology,rainey1998adaptive,martin2013multiple,blount2008historical}? They could also be the result of exploration of new habitats, disturbance opening up niche space to be invaded~\cite{sneppen1995evolution,sole1996extinction}, or something else entirely. Our current analysis cannot answer these questions clearly, but the evidence does clearly show that an explanation is necessary. Second, we have shown that a class of distinct fast processes all map on to the same observable phenomena at coarse temporal resolutions through a combination of their parameters.  This is a quantitative example of a long-discussed idea in ecology that only a handful of parameters survive to describe phenomena at larger or longer scale. The assumption is inherent in neutral models, but also in other, simplified models of macroecological patterns~\cite{mcgill2010towards}. Our approach can form the starting point of quantitatively understanding which parameters and processes `upscale', and which do not. Finally, why do we see such clearly convergent patterns across divergent habitat types?  The ecological and evolutionary constraints leading to the patterns we've seen deserve a fuller explanation.

\section*{Acknowledgments}
A.D.B. was supported by a CompGen Fellowship from the University of Illinois and by the Cooperative State Research, Education, and Extension Service, US Department of Agriculture, under project number ILLU 875-952. J.O.D. acknowledges the Simons Foundation Grant \#376199, McDonnell Foundation Grant \#220020439, NSF \#DEB1557192 and Templeton World Charity Foundation Grant \#TWCF0079/AB47. Sample processing, sequencing and core amplicon data analysis were performed by the Earth Microbiome Project (www.earthmicrobiome.org), and all amplicon sequence data and metadata have been made public through the data portal (qiita.microbio.me/emp).

\section*{Author contributions}
A.D.B. and J.O.D. designed the study and wrote the manuscript. A.D.B. developed analytical tools and analysed phylogenetic data.

\section*{Corresponding author}
Correspondences to Alice Doucet Beaupr\'e (doucetb2@illinois.edu)

\setcounter{section}{0}
\setcounter{figure}{0}

 \clearpage

{\Large
\textbf\newline{Supplementary Information: Widespread bursts of diversification in microbial phylogenies} 
}
\newline

\section{Formalism}
\subsection{Generating functions (GF) and holomorphic/Fock-space formalism}
The equivalence between holomorphic/meromorphic generating functions and Fock-space methods applied to classical objects has a long history in the field of nonequilibrium statistical mechanics of many-body systems, beginning with the seminal papers \cite{doi_second_1976,peliti_path_1985}. Yet it is only relatively recently that those methods have been recognized as potent tools applicable to mathematical biology and ecology\cite{sasai_stochastic_2003,dodd_many-body_2009}. Let the probability generating function (PGF) of the state $\psi$ at time $t$
$$\psi_t(z) = \sum_{n\ge 0}p_n(t) z^n \equiv \sum_{n\ge 0}p_n(t)\vert n\rangle = \vert\psi_t\rangle$$
where $p_n(t)$ represents the probability of having abundance $n$, or more to the point, of having $n$ lineages at time $t$. We adopt the convention that roman letters inside a ket $\vert n\rangle$ are used to denote a monomial/unit mass/abundance state $\vert n\rangle\equiv z^n$ while Greek letters inside a ket $\vert \psi\rangle$ denote mixtures like above. Single abundance state probabilities can be extracted by successive derivation
\begin{equation}
p_n(t) = \left[z^n\right] \psi_t\left(z\right)=\frac{1}{n!}\left.\frac{\partial^n}{\partial z^n}\psi_t(z)\right\vert_{z=0} \equiv \frac{1}{n!}\langle n\vert \psi_t\rangle.
\end{equation}
In the Fock-space formalism  we write the explicit scalar product
\begin{equation}
\langle m\vert n\rangle = m!\delta_{m,n}
\end{equation}
and the completeness relation
\begin{equation}\label{eq:completeness}
\mathbb{1} = \sum_n \frac{\vert n\rangle\langle n\vert}{n!}.
\end{equation}
The normalization of the probability distribution $p_n(t)$ implies that $\psi_t(1) = 1$. Evaluating a GF at a point $z$ requires the introduction of a left coherent-state, denoted with an underline,
\begin{equation}
\psi_t(z) = \langle\underline z\vert\psi_t\rangle = \langle 0\vert e^{z \hat{a}}\vert\psi_t\rangle.
\end{equation}
\textit{En passant} this innocent looking equation is at the core of the equivalence between holomorphic functions and state vectors in Fock spaces. It allows us to easily translate between the natural, combinatorially intuitive language of generating functions, which we will use profusely in the following to construct our coarse-grained conditioned tree observables (or chunks), and the practical linear algebra methods forming the numerical backbone of this study. Under this equivalence the normalization condition becomes $\langle \underline 1\vert\psi_t\rangle = 1$. We use this particular coherent state to find the expectation value of operators, namely
\begin{equation}\langle \hat{\mathcal{O}}\rangle_\psi = \langle\underline{1}\vert\hat{\mathcal{O}}\vert\psi\rangle = \sum_{mn} O_{mn} p_n, \quad O_{mn} = \frac{1}{m!}\langle m\vert\hat{\mathcal{O}}\vert n\rangle.
\end{equation}
Classical stochastic observables are usually diagonal, i.e $O_{mn} = o_{n}\delta_{m,n}$. The holomorphic representation for creation and annihilation operators corresponds to multiplication and derivation by $z$, i.e. $\hat a^\dagger\equiv z$ and $\hat a\equiv \partial_z$, and satisfy bosonic commutation relation
\begin{equation}
\left[\partial_z, z\right] = 1.
\end{equation}
The master equation for a continuous-time stochastic process is written in the language as the partial differential equation (PDE)
\begin{equation}\label{eq:ME}
\partial_t\psi_t(z) = \mathcal{L}\left[z, \partial_z\right]\psi_t(z)
\end{equation}
together with initial condition $\psi_0(z)$. $\mathcal{L}[z, \partial_z]$ is the time evolution generator and encodes all the information about the instantaneous dynamics of the process. Brackets here denote the dependence of the generator on $z$ and $\partial_z$, not its multiplication by the bosonic commutator. Eq.~(\ref{eq:ME}) admits the formal solution
\begin{equation}
\psi_t(z) = e^{\mathcal{L}[z, \partial_z]t}\psi_0(z) \equiv \langle\underline z\vert e^{\mathcal{L}[\hat a^\dagger, \hat a] t}\vert\psi_0\rangle.
\end{equation}
This is the starting point for the numerical exponentiation scheme used below and in the main text to construct the likelihood of a coarse-grained (CG) tree.

The holomorphic formalism admits three important similarity transformations\cite{mattis_uses_1998}: two shifts,
\begin{equation}\label{eq:shiftz}
e^{x\partial_z}\psi(z, \partial_z)e^{-x\partial_z} = \psi(z + x, \partial_z)
\end{equation}
and
\begin{equation}\label{eq:shiftdz}
e^{x z}\psi(z, \partial_z)e^{-x z} = \psi(z, \partial_z - x),
\end{equation}
and a scaling transformation,
\begin{equation}\label{eq:scale}
e^{x z\partial_z} \psi(z, \partial_z) e^{-x z\partial z} = \psi(z e^x, e^{-x}\partial_z).
\end{equation}
\subsection{Incomplete lineage sampling}
If we approximate the sampling process by a Bernouilli trial with success probability, or sampling fraction, $f$, then the GF for the joint probability of successfully sampling $n$ individuals out of a population of $n$ or more individuals is given by
\begin{equation}
\psi_t(1 - f + fz) = \sum_{n\ge0} p_n(t)(1 - f + f z)^n = \sum_{n\ge 0}\left(\sum_{k\ge 0} p_{n+k}(t)\binom{n+k}{k}(1-f)^k\right)f^n z^n.
\end{equation}
This expression captures the fact that every states with abundance greater than $n$ contribute to the probability of sampling exactly $n$ lineages.

\section{Processes}
\subsection{Birth process}
The birth process with per capita birth rate $b$ consists in the transition
$$A\overset{b}{\longrightarrow} 2A$$
The generator for the birth process is given by
\begin{equation}\label{eq:birthgenerator}
\mathcal{L}_B = b(\hat a^\dagger - 1)\adag \hat{a}.
\end{equation}
The instantaneous transition rates between states are given by
\begin{equation}\begin{split}
L_{B,mn} &= \frac{1}{m!}\langle m\vert \mathcal{L}_B\vert n\rangle,\\
&= \frac{b}{m!}\left(n \langle m\vert n+1\rangle - n \langle m\vert n\rangle\right),\\
&= \frac{bn}{m!}\left(m!\delta_{m,n+1} - m!\delta_{m,n}\right),\\
&= bn\left(\delta_{m,n+1} - \delta_{m,n}\right).
\end{split}\end{equation}
The generator, in matrix form,
\begin{equation}\label{eq:LB}
\hat{L}_B = \left(
\begin{array}{crrrrc}
0 & 0 & 0 & 0 & 0 & \ldots \\
0 & -b & 0 & 0 & 0 & \\
0 & b & -2b & 0 & 0 & \\
0 & 0 & 2b & -3b & 0 & \\
0 & 0 & 0 & 3b & -4b & \\
\vdots & & & & & \ddots
\end{array}
\right).
\end{equation}
\subsection{Death process}
The death process with per capita death rate $d$ consists in the transition
$$ A\overset{d}{\longrightarrow} \emptyset.$$
The generator of the death process is given by
\begin{equation}
\mathcal{L}_D = d\left(1 - \adag\right)\hat{a}
\end{equation}
and the instantaneous transition rates
\begin{equation}
L_{D,mn} = dn\left(\delta_{m+1, n} - \delta_{m,n}\right)
\end{equation}
\begin{equation}\label{eq:LD}\Rightarrow \hat{L}_D = \left(
\begin{array}{crrrrc}
0 & d & 0 & 0 & 0 & \ldots \\
0 & -d & 2d & 0 & 0 & \\
0 & 0 & -2d & 3d & 0 & \\
0 & 0 & 0 & -3d & 4d & \\
0 & 0 & 0 & 0 & -4d & \\
\vdots & & & & & \ddots
\end{array}
\right).\end{equation}
\subsection{Innovation process}
The innovation process consists in initiating at per capita rate $\rho$ a Yule (pure birth) process and identifying all finite-time transition probabilities with infinitesimal transition rates. To obtain the innovation process generator we first need to solve the birth process exactly. We begin by writing its formal solution using Eqs.~\ref{eq:birthgenerator} and \ref{eq:ME} as
\begin{equation}
\psi_t(z) = e^{\sigma\tau(z - 1)z\partial z}\psi_0(z).
\end{equation}
Using the change of variable $y = 1/z$ we rewrite the evolution equation
\begin{equation}
\psi_\tau\left(1/y\right) = e^{\sigma\tau (y - 1)\partial y}\psi_0\left(1/y\right).
\end{equation}
Using Eqs.~\ref{eq:shiftz} and \ref{eq:scale} we solve
\begin{equation}\begin{split}
\psi_\mathcal{T}\left(1/y\right) &= e^{-\partial_y}e^{\sigma\tau y\partial_y}e^{\partial_y}\psi_0(1/y),\\
&=\psi_0\left(\frac{1}{(y - 1)e^{\sigma\tau} + 1}\right),\\
&= \psi_0\left(\frac{(1-\alpha)z}{1- \alpha z}\right),
\end{split}\end{equation}
where $\alpha = 1 - e^{-\sigma\tau}$. For a given burst time-scale $\tau$, higher fitness processes lead to values of $\alpha$ closer to 1. For a single initial lineage, $\psi_0(z) = z$ and we recover the geometric PGF
\begin{equation}
\psi_\tau(z) = \frac{(1-\alpha)z}{1-\alpha z} = (1-\alpha)\sum_{n\ge 1}\alpha^{n-1} z^n.
\end{equation}
The generator of the innovation process
$$A \overset{\rho g_\alpha(n)}{\longrightarrow} nA,$$
where $g_\alpha(n) = (1-\alpha)\alpha^{n-1}$, is therefore given by
\begin{equation}\begin{split}
\mathcal{L}_I &= \rho\left(\frac{(1-\alpha)\adag}{1-\alpha \adag} - \adag\right)\hat{a},\\
&= \rho\alpha\left(\frac{(1-\alpha)\adag}{1-\alpha\adag} - 1\right)\adag\ha,\\
&= \rho\alpha\frac{(\adag-1)}{1-\alpha\adag}\adag\ha.
\end{split}\end{equation}
In the second equality, we made sure to only include off-diagonal transitions $1\rightarrow k\ge 2$ in the positive term by absorbing the false transition $1\rightarrow 1$ in the negative diagonal term. Doing so also highlights the fact that the effective innovation rate is actually equal to $\rho\alpha$ rather than $\rho$. It is tempting to interpret, in line with the Red Queen hypothesis, the rate $\rho(1-\alpha)$ of initiations of ``invisible" innovations as capturing perhaps the rate at which lineages must constantly innovate in short bursts of higher fitness just to keep pace with co-occurring lineages and changing environments.
The instantaneous transition rates are given by
\begin{equation}
L_{I, mn} = \rho n(1-\alpha)\alpha^{m-n}\delta_{m-n\ge 1} - \rho\alpha n\delta_{m,n}
\end{equation}
and the matrix of the generator
\begin{equation}\label{eq:LI}\hat{L}_I = \left(
\begin{array}{cccccc}
0 & 0 & 0 & 0 & 0 & \ldots \\
0 & -\rho\alpha & 0 & 0 & 0 & \\
0 & \rho(1-\alpha)\alpha & -2\rho\alpha & 0 & 0 & \\
0 & \rho(1-\alpha)\alpha^2 & 2\rho(1-\alpha)\alpha & -3\rho\alpha & 0 & \\
0 & \rho(1-\alpha)\alpha^3 & 2\rho(1-\alpha)\alpha^2 & 3\rho(1-\alpha)\alpha & -4\rho\alpha & \\
\vdots & & & & & \ddots
\end{array}
\right).
\end{equation}
\subsection{Heterogeneous innovation process}
To account to the possibility of multiple innovation processes with different values of $\alpha$ we compound the parameter of the geometric part with its conjugate prior the beta distribution. The generator
\begin{equation}\begin{split}
\mathcal{L}_H &= \eta \int_0^1 \text{Beta}(x\vert\alpha, \beta)x\left(\frac{(1-x)\adag}{1-x\adag} - 1\right)\adag\ha dx,\\
&= \eta \frac{\alpha}{\alpha + \beta}\left(\frac{\beta}{\alpha + \beta +1}{}_2 F_1\left(1, \alpha + 1; \alpha + \beta + 2; \adag\right) \adag - 1\right)\adag\ha\\
&= \eta \overline m\left(\frac{s(1-\overline m)}{s + 1}{}_2 F_1\left(1, s\overline m + 1; s + 2; \adag\right)\adag - 1\right)\adag \ha,\\
&=\eta\overline{m}{}_2F_1\left(1, s\overline{m}+1, s + 1; \adag\right)(\adag - 1)\adag\ha.
\end{split}\end{equation}
In the last two lines we used the reparametrization of the beta distribution in terms of the precision $s = \alpha + \beta$, $s > 0$, and the mean $\overline m = \alpha/(\alpha + \beta)$, $0 < \overline m < 1$. Similarly as before the instantaneous rates
\begin{equation}
\begin{split}
L_{H,mn} &= \eta n \frac{\beta \Gamma(\alpha + \beta)\Gamma(\alpha + m - n)}{\Gamma(\alpha)\Gamma(\alpha + \beta + 1 + m - n)}\delta_{m-n\ge 1} - \eta n \frac{\alpha}{\alpha + \beta}\delta_{m,n}\\
&= \eta n\frac{s(1-\overline m)\Gamma(s)\Gamma(s\overline m + m - n)}{\Gamma(s\overline m)\Gamma(s + 1 + m - n)}\delta_{m-n\ge 1} - \eta n \overline m \delta_{m,n}.
\end{split}
\end{equation}
For $k=m-n\gg 1$, instantaneous rates follow the power law
\begin{equation}
L_{H,mn} \sim \frac{\Gamma(s\overline{m} + k)}{\Gamma(s + 1 + k)} \sim \frac{1}{k^{s(1-m)+1}} = \frac{1}{k^{\beta + 1}}.
\end{equation}
The power law phase is controlled by the $\beta$ parameter of the beta distribution. The parameter $\beta$ in turn controls the shape of the density of the geometric innovation parameters around $x=1-\epsilon$. As $\epsilon\ll 1$ approaches 0 and $x$ approaches 1, the geometric distribution acquires a progressively longer exponential decay that tends towards a uniform improper distribution over all $k\ge 1$. When $\beta < 1$, the density diverge algebraically as $\epsilon^{\beta - 1}$ and the compounding of wide geometric distributions give rise to power laws with tail exponent between 1 and 2. When $\beta = 1$, then the tail exponent becomes exactly 2, and when $\beta > 1$, the tail exponent becomes greater than 2. Finally the matrix of the generator
\begin{equation}\label{eq:LH}
\hat{L}_H = \left(
\begin{array}{cccccc}
0 & 0 & 0 & 0 & 0 & \ldots \\
0 & -\eta\overline m & 0 & 0 & 0 & \\
0 & \eta \frac{s(1-\overline m)\Gamma(s)\Gamma(s\overline m + 1)}{\Gamma(s\overline m)\Gamma(s + 2)} & -2\eta\overline m & 0 & 0 & \\
0 & \eta\frac{s(1-\overline m)\Gamma(s)\Gamma(s\overline m + 2)}{\Gamma(s\overline m)\Gamma(s + 3)} & 2\eta\frac{s(1-\overline m)\Gamma(s)\Gamma(s\overline m + 1)}{\Gamma(s\overline m)\Gamma(s + 2)} & -3\eta\overline m & 0 & \\
0 & \eta\frac{s(1-\overline m)\Gamma(s)\Gamma(s\overline m + 3)}{\Gamma(s\overline m)\Gamma(s + 4)} & 2\eta\frac{s(1-\overline m)\Gamma(s)\Gamma(s\overline m + 2)}{\Gamma(s\overline m)\Gamma(s + 3)} & 3\eta\frac{s(1-\overline m)\Gamma(s)\Gamma(s\overline m + 1)}{\Gamma(s\overline m)\Gamma(s + 2)} & -4\eta\overline m & \\
\vdots & & & & & \ddots
\end{array}
\right).
\end{equation}

\section{Maximum Likelihood approach}
\subsection{CG tree observables and likelihood}
We extract dynamical information contained in a chronogram (ultrametric phylogenetic tree) by looking at CG transitions, i.e. single lineages going to many lineages at a time closer to the present. Those are called chunks in the main text and to each chunk is associated a tuple $(t, s, k)$ and likelihood chunk $\phi_f^{(k)}(t, s)$ where $0<f\le 1$ is the sampling fraction.  A likelihood chunk represents the probability that a unique lineage at time $t$ in the past had exactly $k\ge 1$ descendants at time $s \le t$ that each had at least 1 extant lineage in the present, while all other lineages above $k$ were unobserved by virtue of having gone extinct or of being missed during sampling of intensity $f$. We can write using the previous formalisms
\begin{equation}\begin{split}\label{eq:chunk}
\phi_f^{(k)}(t, s\vert\theta) &= \frac{1}{k!}\frac{\left(1 - \mathcal{U}_s(1-f\vert\theta)\right)^k \mathcal{U}^{(k)}_{t-s}(\mathcal{U}_s(1-f\vert\theta)\vert\theta)}{1 - \mathcal{U}_t(1-f\vert\theta)},\\
&= \frac{1}{k!}\frac{\left(1 - \langle\underline{1-f}\vert e^{\mathcal{L(\theta)} s}\vert 1\rangle\right)^k \langle\underline{1-f}\vert e^{\mathcal{L(\theta)}s}\hat{a}^ke^{\mathcal L(\theta)(t-s)}\vert 1\rangle}{1 - \langle\underline{1-f}\vert e^{\mathcal{L(\theta)} t}\vert 1\rangle},
\end{split}\end{equation}
where $\mathcal{U}_t(z\vert\theta)$ is the solution of Eq.~\ref{eq:ME} with initial condition $\mathcal{U}_0(z\vert\theta) = z$ and $\mathcal{U}^{(k)}_t(z\vert\theta)$ is its $k$-th derivative w.r.t. $z$. 
In terms of probabilities contained in $\mathcal U_t(z\vert\theta) = \sum_{k\ge 0} P(k\vert 1, t,\theta)z^k$ and $\left(\mathcal U_t(z\vert\theta)\right)^m = \sum_{k\ge 0} P(k\vert m, t, \theta)z^k$ the expression \ref{eq:chunk} for chunks is given by
\begin{equation}\begin{split}\label{eq:chunkP}
\phi_f^{(k)}(t,s\vert\theta) &= \frac{\sum_{m,n = 0}^\infty (1-f)^m P(m\vert n, s,\theta) {n+k\choose k}P(n+k\vert 1,  t - s,\theta)}{1 -\sum_{m=0}^\infty (1-f)^mP(m\vert 1, t,\theta)} \\
& \times \left(1 - \sum_{m=0}^\infty (1-f)^m P(m\vert 1,s,\theta)\right)^k.
\end{split}\end{equation}
This is a probability distribution over $k\ge 1$. To show this we sum over $k$
\begin{equation}
\begin{split}
\sum_{k\ge 1}\phi^{(k)}_f(t, s\vert \theta) &= \frac{\sum_{k\ge 0}\left. \frac{(1 - \mathcal{U}_s(1-f))^k}{k!}\partial^k_z \mathcal{U}_{t-s}(z)\right\vert_{z=\mathcal{U}_s(1-f)} - \mathcal{U}_{t-s}(\mathcal{U}_s(1-f))}{1 - \mathcal{U}_t(1-f)},\\
&= \frac{\left. e^{(1-\mathcal{U}_s(1-f))\partial_z}\mathcal{U}_{t-s}(z)\right\vert_{z=\mathcal{U}_s(1-f)} - \mathcal{U}_t(1 - f)}{1 - \mathcal{U}_t(1-f)},\\
&= \frac{\mathcal{U}_{t-s}(\mathcal{U}_s(1-f) + 1 - \mathcal{U}_s(1-f)) - \mathcal{U}_t(1-f)}{1 - \mathcal{U}_t(1-f)},\\
&= \frac{\mathcal{U}_{t-s}(1) - \mathcal{U}_t(1-f)}{1 - \mathcal{U}_t(1-f)},\\
&= 1.
\end{split}
\end{equation}
In the first equality we added and subtracted the extinction term $k=0$. In the second equality we used the semigroup/Chapman-Kolmogorov property. In the third equality we used the shift property Eq.~\ref{eq:shiftz}. In the fourth equality we cancelled terms inside the argument of $\mathcal{U}_{t-s}$ and finally used the normalization condition $\mathcal{U}_{t-s}(1) = 1$. All terms are positive and sum to one and therefore $\phi^{(k)}_f(t, s\vert\theta)$ is a probability distribution over $k\ge 1$.

Combining the method of characteristics and numerical complex derivation, the first (holomorphic) formulation of Eq.~\ref{eq:chunk} allows for the exact numerical computation of chunk likelihoods. Yet we found the second (Fock space) formulation using truncated matrix exponentials and linear algebra to be easier to implement and more stable at large values of $k$. All the information of a given model reside inside the generator $\mathcal L(\theta)$. The Greek letter $\theta$ is understood as the set of all parameters of a given model. We will denote the action of CGing/slicing with slice width $\Delta=T/K$ a chronogram $\mathcal T$ by
\begin{equation}
\mathcal R_K[\mathcal{T}] = \left\lbrace (t_i, s_i, k_i)\right\rbrace_{i\in I}.
\end{equation}
The index set $I$ runs over chunk indices following some traversal order over the CG chronogram. We can write the likelihood of a given CG chronogram
\begin{equation}\label{eq:cgprob}
\text{Pr}\left[\mathcal R_K[\mathcal{T}]\vert\theta, f\right] = \prod_{i\in I} \phi_f^{(k_i)}(t_i, s_i\vert \theta),\quad t_i - s_i = \Delta \forall i\in I.
\end{equation}
The central inference objective of this framework is to seek ML estimates (MLE) $\theta^*$ by maximizing the log-likelihood
\begin{equation}\label{eq:mle}
\theta^* = \arg\max_\theta \log\text{Pr}\left[\mathcal R_K[\mathcal{T}]\vert\theta,f\right] = \arg\max_\theta \sum_{i\in I} \log\phi_f^{(k_i)}(t_i, s_i, k_i\vert\theta).
\end{equation}
\subsection{Equivalence with the Morlon \textit{et al.} likelihood}
It is worth mentioning here that Eqs.~\ref{eq:chunk} and \ref{eq:cgprob} recover the likelihood found in \cite{Morlon2011fossil} in the limit of infinitesimal chunk duration $t-s = \delta t$. This limit is also equivalent to a particular slicing scheme where each branch gets a chunk of size $k=1$ with duration $t-s$ equal the length of the branch, and each node a chunk of size $k=2$ realized with an infinitesimal duration. Let us first show the equivalence between the infinitesimal and branch-node slicing. Consider two consecutive branch segments of length $t - s$ and  $s - r$. Together they contribute
\begin{equation}\label{eq:weight2}
\phi^{(1)}_f(t,s)\phi^{(1)}_f(s,r) =\left[\frac{\mathcal{U}_{t-s}^{(1)}(\mathcal{U}_s(1-f))}{1-\mathcal{U}_t(1-f)}(1-\mathcal{U}_s(1-f))\right]\left[\frac{\mathcal{U}_{s-r}^{(1)}(\mathcal{U}_r(1-f))}{1-\mathcal{U}_s(1-f)}(1-\mathcal{U}_r(1-f))\right]
\end{equation}
to the full tree likelihood. Recall the Chapman-Kolmogorov semi-group property of generating functions which ensures $\mathcal{U}_{t-s}(\mathcal{U}_{s-r}(z)) = \mathcal{U}_{t-r}(z)$. Therefore
\begin{equation}\begin{split}
\mathcal{U}^{(1)}_{t-r}(\mathcal{U}_r(1-f)) &= \left.\frac{\partial}{\partial z}\mathcal{U}_{t-s}(\mathcal{U}_{s-r}(z))\right\vert_{z=\mathcal{U}_r(1-f)},\\
&= \mathcal{U}^{(1)}_{t-s}(\mathcal{U}_{s-r}(\mathcal{U}_r(1-f)))\mathcal{U}^{(1)}_{s-r}(\mathcal{U}_r(1-f)),\\
&= \mathcal{U}^{(1)}_{t-s}(\mathcal{U}_s(1-f)) \mathcal{U}^{(1)}_{s-r}(\mathcal{U}_r(1-f)).
\end{split}\end{equation}
Substituting backward in Eq.~\ref{eq:weight2} it follows that
\begin{equation}\begin{split}\label{eq:11composition}
\phi^{(1)}_f(t,s)\phi^{(1)}_f(s,r) &= \frac{1-\mathcal{U}_s(1-f)}{1-\mathcal{U}_t(1-f)}\frac{1-\mathcal{U}_r(1-f)}{1-\mathcal{U}_s(1-f)}\left[\mathcal{U}_{t-s}^{(1)}(\mathcal{U}_s(1-f))\mathcal{U}_{s-r}^{(1)}(\mathcal{U}_r(1-f))\right],\\
&= \frac{\mathcal{U}_{t-r}(\mathcal{U}_r(1-f))}{1-\mathcal{U}_t(1-f)}(1-\mathcal{U}_r(1-f)),\\
&= \phi^{(1)}_f(t, r).
\end{split}\end{equation}
This means that the product of consecutive chunks of size $k=1$ compounds into one chunk of size $k=1$ with length equal to the sum of individual chunk lengths. This in turn implies the equivalence between the infinitesimal slicing of a tree and a scheme where each branch $i$ gets its own $k=1$ chunk with the same initial and final times $t_i$ and $s_i+\delta t$, while its immediate downstream node inherits a chunk of size $k=2$ with initial and final times $s_i+\delta t$ and $s_i$. 
Chunks ``touching the present" contribute a weight
\begin{equation}\begin{split}
\phi^{(k)}_f(t, 0) &= \frac{1}{k!}\frac{\mathcal{U}_{t}^{(k)}(\mathcal{U}_0(1-f))}{1-\mathcal{U}_t(1-f)}(1-\mathcal{U}_0(1-f))^k,\\
&= \frac{f^k}{k!}\frac{\mathcal{U}_t^{(k)}(1-f)}{1-\mathcal{U}_t(1-f)}.
\end{split}\end{equation}
Finally node contributions for the BD model in this scheme
\begin{equation}\begin{split}
\phi^{(2)}_f(s+\delta t, s)&= \frac{1}{2}\frac{\mathcal{U}_{\delta t}^{(2)}(U_s(1-f))}{1-\mathcal{U}_{s+\delta t}(1-f)}(1-\mathcal{U}_s(1-f))^2,\\
&\sim b\delta t \frac{(1 - \mathcal{U}_s(1-f))^2}{1-U_{s+\delta t}(1-f)}.
\end{split}\end{equation}
Therefore the contribution of an internal branch starting at $t$ in the past and ending with a node at $s$ has weight
\begin{equation}\begin{split}
\phi^{(1)}_f(t, s+\delta t)\phi^{(2)}_f(s+\delta t, s) &= \frac{\mathcal{U}_{t-s-\delta t}^{(1)}(\mathcal{U}_{s +\delta t}(1-f))}{1- \mathcal{U}_t(1-f)}\frac{1 - \mathcal{U}_{s+\delta t}(1-f)}{1 - \mathcal{U}_{s+\delta t}(1-f)} (1 - \mathcal{U}_s(1-f))^2 b\delta t,\\
&= b\delta t \frac{\mathcal{U}_{t-s}^{(1)}(\mathcal{U}_{s}(1-f))}{1- \mathcal{U}_t(1-f)}(1 - \mathcal{U}_s(1-f))^2,
\end{split}\end{equation}
and the tree likelihood in the limit $K\rightarrow\infty$ for a tree of depth $t$ and $n$ leaves
\begin{equation}
\text{Pr}[\mathcal{T}_\infty] = \frac{\left[\prod_{j\in\text{int.branches}} b \delta t \mathcal{U}^{(1)}_{t_j-s_j}(\mathcal{U}_{s_j}(1-f))\right]\left[\prod_{i\in\text{leaves}}\mathcal{U}^{(1)}_{t_i}(1-f)\right]}{1-\mathcal{U}_t(1-f)}.
\end{equation}
This is equivalent to Eq.~(1) in \cite{Morlon2011fossil} for the case of constant birth and death rates modulo a factor $\delta t^{(n-1)}$. Indeed for the BD model
\begin{equation}
\mathcal U_t(z) = p_0(t) + (1- p_0(t))\frac{\left(1 - \frac{p_0(t)}{r}\right)z}{1 - \frac{p_0(t)}{r}z}
\end{equation}
where the absorbing state/exctinction probability
\begin{equation}
p_0(t) = r\frac{\omega(t)-1}{\omega(t) - r},\qquad \omega(t) = e^{b(1-r)t},
\end{equation}
and $r= d/b$ the ratio of constant per capita death and birth rates. We already found the Yule limit previously when we constructed the innovation process. In the Yule limit  $r\rightarrow 0$, therefore $p_0\rightarrow 0$ and $\alpha = p_0(t)/r\rightarrow 1-e^{-bt}$. Generalizing to time-varying rates is straightforward and the equivalence remains valid.
\subsection{Chunk generating function}
Using manipulations similar to those of the previous sections, we find the chunk PGF
\begin{equation}
\begin{split}
\tilde{\Phi}_f(t,s,y) =& \sum_{k\ge 0} \frac{y^k}{k!} \mathcal U_{t-s}^{(k)}\left(\mathcal U_s(1-f)\right)\left(1-\mathcal U_s(1-f)\right)^k,\\
&= \left. e^{y(1-\mathcal U_s(1-f))\partial_z}\mathcal U_{t-s}(z)\right\vert_{z=\mathcal U_s(1-f)},\\
&= \mathcal U_{t-s}\left(\mathcal U_s(1-f) + y\left(1 - \mathcal U_s(1-f)\right)\right).
\end{split}
\end{equation}
This expression has a very intuitive combinatorial interpretation that could have been guessed from the beginning and used as the starting point for the chunk decomposition. It symbolically stipulates that at $t-s$, for whichever state you are in, substitute each lineage represented by atoms $z$ with a Bernouilli trial of success probability equal to the probability of surviving across time $s$ to the present. In other words, lineages are "preemptively" split at time $t-s$ into empty atoms (the monomial 1) weighted by the probability of not making it to the present, $\mathcal U_s(1-f)$, and unit atoms (the monomial $y$) weighted by the probability of making it to the present, $1 - \mathcal U_s(1 - f)$. We easily recover the extant chunk PGF when conditioning on survival by omitting the extinction term at $y=0$ and renormalizing. We find
\begin{equation}\begin{split}\label{eq:chunkgf}
\Phi_f(t, s, y) &= \frac{\tilde\Phi_f(t, s, y) - \tilde\Phi_f(t, s, 0)}{1 - \tilde\Phi_f(t, s, 0)},\\
&= \frac{\mathcal U_{t-s}\left(\mathcal U_s(1-f) + y\left(1 - \mathcal U_s(1-f)\right)\right) - \mathcal U_t(1-f)}{1 - \mathcal U_t(1-f)}
\end{split}\end{equation}
and finally it is straightforward, if tedious, to verify that
\begin{equation}
\phi^{(k)}_f(t, s) = \frac{1}{k!}\left.\frac{\partial^k}{\partial y^k}\Phi_f(t, s, y)\right\vert_{y=0},\quad k\ge 1,
\end{equation}
recovers Eq.~\ref{eq:chunk}.

The chunk generating function also satisfy the Chapman-Kolmogorov property
\begin{equation}
\Phi(t, s, z) = \Phi(t, r, \Phi(r, s, z)),
\end{equation}
which we can verify using Eq.~\ref{eq:chunkgf} and the Chapman-Kolmogorov property for $\mathcal{U}_t(z)$. This property of the chunk generating function translates at the level of likelihood chunks thus:
\begin{equation}\begin{split}\label{eq:chunkbreakup}
\phi^{(k)}_f(t, s) &= \sum_{c\in\text{Comp}(k)} \phi^{(\vert c\vert)}_f(t,r)\prod_{\lambda \in c}\phi^{(\lambda)}_f(r, s),\\
&= \sum_{\pi\in\text{Part}(k)} \vert\text{Comp}(\pi)\vert\phi^{(\vert c\vert)}_f(t,r)\prod_{\lambda \in \pi}\phi^{(\lambda)}_f(r, s),\\
\end{split}\end{equation}
where $\text{Comp}(k)$ and $\text{Comp}(\pi)$ are respectively the set of compositions of the integer $k$ and by a slight abuse of notation the set of compositions equivalent to an integer partition $\pi$, i.e. a composition is equivalent to a partition iff both multisets of their parts are equal. E.g. $(1, 5, 4, 5) \sim (5, 5, 4, 1)$, but $(1, 5, 4, 5)\not\sim(6, 5, 4, 2)$. $\text{Part}(k)$ represents the set of partitions of the integer $k$. Finally for $\pi$ a partition of $k$, the size of the set of its equivalent compositions is given by the multinomial coefficient
\begin{equation}
\vert\text{Comp}(\pi)\vert = {\vert \pi\vert \choose m_\pi(1), m_\pi(2), \dots, m_\pi(k)} = \frac{\vert \pi\vert !}{\prod_{i=1}^k m_\pi(i)!},
\end{equation}
with $m_\pi(i)$ the multiplicity of parts of size $i$ in $\pi$. Eq.~\ref{eq:chunkbreakup} describes how to breakup a chunk into a sum-product of chunks of smaller or equal size, effectively enumerating elements of its equivalence class. Applying this decomposition recursively one can see how a coarse-grained phylogeny is equivalently indexed by a set of time slices (the $r$'s so to speak) together with an integer composition for each slice. Moreover it naturally gives rise to the Hastings ratio used in the Metropolis-Hastings MCMC algorithm described in Section~\ref{subseq:mcmc} where coarse-grained phylogenies are sampled using the space of integer multipartitions.
\subsection{Combined models, nestedness, and likelihood ratio test (LRT)}
The Birth-Death model (BD) is obtained by adding the Birth/Yule (B) and Death (D) processes together to obtain the BD generator
\begin{equation}
\hat L_{BD}(\theta) = \hat L_{B}(b) + \hat L_{D}(d), \quad\theta = \lbrace b, d\rbrace.
\end{equation}
Similarly for the Birth-Death-Innovation (BDI) model
\begin{equation}
\hat L_{BDI}(\theta) = \hat L_{BD}(b, d) + \hat L_{I}(\rho, \alpha), \quad\theta = \lbrace b, d, \rho, \alpha\rbrace,
\end{equation}
and the Birth-Death-Heterogeneous innovation (BDH) model
\begin{equation}
\hat L_{BDH}(\theta) = \hat L_{BD}(b, d) + \hat L_{H}(\eta, s, \overline{m}), \quad\theta = \lbrace b, d,\eta, s, \overline m\rbrace.
\end{equation}
All three models are nested. BD is recovered from BDI in the limit $\rho\alpha\rightarrow 0$, and BDI is recovered from BDH in the limit $\overline{m}\rightarrow \alpha$,  $s\rightarrow \infty$, and $\eta\rightarrow \rho$. We can therefore perform model selection using the LRT, i.e. comparing the statistic
\begin{equation}\label{eq:LRTD}D = 2\left(\log\text{Pr}\left[\mathcal R_K[\mathcal{T}]\vert\theta_{\text{alternative}}\right] - \log\text{Pr}\left[\mathcal R_K[\mathcal{T}]\vert\theta_{\text{null}}\right]\right)
\end{equation}
against the $\chi^2_{\text{ddof}}$ distribution with $\text{ddof}=\vert\theta_{\text{alternative}}\vert - \vert\theta_{\text{null}}\vert$ degrees of freedom. $\text{ddof} = 2$ for BD vs BDI, and $\text{ddof}=1$ for BDI vs BDH.
\subsection{Numerical exponentiation and truncation}
Optimizing Eq.~\ref{eq:mle} requires the evaluation of Eq.~\ref{eq:chunk}, equivalently Eq.~\ref{eq:chunkP}, for arbitrary values of $\theta$. To do so we use the completeness relation Eq.~\ref{eq:completeness} to transform Eq.~\ref{eq:chunk} into an explicit matrix-vector multiplication, e.g.
\begin{equation}\begin{split}
\langle\underline{1-f}\vert &\mathbb{1}e^{\mathcal{L(\theta)}s}\mathbb{1}\hat{a}^k\mathbb{1}e^{\mathcal L(\theta)(t-s)}\mathbb{1}\vert 1\rangle =\\ &\sum_{m,n,p,q}\langle\underline{1-f}\vert m\rangle\frac{1}{m!}\langle m \vert e^{\mathcal L(\theta)s}\vert n\rangle\frac{1}{n!}\langle n\vert \frac{\ha^k}{k!}\vert p\rangle\frac{1}{p!}\langle p\vert e^{\mathcal L(\theta)(t - s)}\vert q\rangle\frac{1}{q!}\langle q\vert 1\rangle.
\end{split}\end{equation}
The vector elements of the left- and right-most bra and kets are
\begin{equation}
\langle\underline{1-f}\vert m\rangle = \sum_{m'} \frac{(1-f)^{m'}\langle m'\vert m\rangle}{m!} = (1-f)^m,
\end{equation}
and
\begin{equation}
\frac{1}{q!}\langle q\vert 1\rangle = \delta_{q,1}.
\end{equation}
The matrix elements of the combination operator
\begin{equation}
\frac{1}{n! k!}\langle n\vert \ha^k \vert p\rangle = \binom{n+k}{k}\delta_{n+k, p} := C^{(k)}_{np}.
\end{equation}
The elements of the two matrix exponentials,
\begin{equation}
\frac{1}{m!}\langle m \vert e^{\mathcal L(\theta)s}\vert n\rangle = \left[e^{\hat L(\theta) s}\right]_{mn},\
\end{equation}
and
\begin{equation}
\frac{1}{p!}\langle p\vert e^{\hat L(\theta)(t - s)}\vert q\rangle = \left[e^{\hat L(\theta)(t - s)}\right]_{pq},
\end{equation}
are obtained using Scipy's $\mathtt{linalg.expm}$ function\cite{al-mohy_new_2009,jones_scipy_2001} and truncated combinations of Eq.~\ref{eq:LB}, \ref{eq:LD}, \ref{eq:LI}, and \ref{eq:LH}. Since those generator matrices are formally of infinite dimension we need an approximation in order to perform any kind of numerics. Similarly to the finite state projection (FSP) algorithm\cite{munsky_finite_2006}, but even more simply, we truncate by taking the $N\times N$ submatrix of $\hat L$ which includes transitions amongst states $0 \le m, n \le N-1$. We omit to correct the diagonal terms and therefore do not compensate for the missing probability flow into states $n \ge N$. This approximation makes $\hat L$, and by extension $e^{\hat Ls}$, non-stochastic and would eventually leak all the probability mass of the initial state. We found this approximation to nonetheless work very well in supercritical cases, e.g. $b > d$ in the BD model, as long as $N$ is chosen at least between $2\times$ or $3\times$ the largest chunk size $\max_{i\in I} k_i$. Almost all maximum likelihood (ML) inference fall inside the supercritical region as expected from the simple fact that most phylogenies grow. It also avoids the build up of probability mass into higher abundances state which is caused by the reflecting boundary inherent to a stochastic truncation (i.e. with corrected diagonal terms). In any case we found results to be insensitive to the choice of leaky vs. non-leaky approximations. Finally we put all pieces together and find for a given chunk
\begin{equation}\begin{split}\label{eq:chunkprod}
\log\phi^{(k)}_f(t, s\vert\theta) &= k\log\left(1 - \sum_m (1-f)^m\left[e^{s\hat L(\theta)}\right]_{m1}\right) - \log\left(1 - \sum_m (1-f)^m\left[e^{t\hat L(\theta)}\right]_{m1}\right) \\
& + \log\sum_{m,n,p} (1-f)^m \left[e^{s\hat L(\theta)}\right]_{mn}C^{(k)}_{np}\left[e^{(t-s)\hat L(\theta)}\right]_{p1},
\end{split}\end{equation}
where all sums are now between $0$ and $N-1$.
\subsection{Full Likelihood construction}
In order to accelerate the evaluation of the complete log-likelihood for a CG phylogeny given by Eq.~\ref{eq:mle} we organize the construction of the likelihood around three observations:
\begin{itemize}
\item all chunks have identical $t - s = \Delta$ and therefore share the same column-vector elements $[e^{(t-s)\hat L(\theta)}]_{m1}$,
\item all chunks in a given slice share row-vector elements for both $\sum_m \left(1-f\right)^m [e^{s\hat L(\theta)}]_{mn}$ and $\sum_m \left(1-f\right)^m [e^{t\hat L(\theta)}]_{mn}$,
\item and slices share interfaces such that, going from the past to the present, one slice's $s_i$ is the next slice's $t_i$, leading to cancellation of all first two terms of the form $\log(1-x)$ in Eq.~\ref{eq:chunkprod} except at the root and leaves.
\end{itemize}
Thus memoization of select function calls or explicit storage of certain parts of the above expression greatly reduces the number of matrix exponentiations and dot products necessary to complete one evaluation of the likelihood. We used Scipy's $\mathtt{optimize.minimize}$ function with the L-BFGS-B\cite{byrd_limited_1995}\blank{\cite{byrd_limited_1995,zhu_algorithm_1997,morales_remark_2011}} algorithm to perform the optimization in Eq.~\ref{eq:mle}. We typically used 5 restarts, each one initiated at the best out of 10 random starting points chosen randomly from a centered normal distribution with variance 2 in transformed coordinates $\log b, \log d, \log \rho, \log \eta, \log s$, and $\logit~\overline m$. We also assume an approximate complete sampling of lineages and set $f=1$.
\section{Goodness of fit test}
\subsection{Exact goodness-of-fit (gof) test}
The LRT allows us to compare two models together. We would also like to assess the quality of the MLE without reference to an alternative model. This is the purpose of the gof. For this purpose we choose the $G$-statistic characterizing the information divergence between empirical and theoretical chunk frequencies. For a CG phylogeny $\mathcal{T}_K = \mathcal R_K[\mathcal{T}]$,
\begin{equation}
G(\mathcal{T}_K) = 2 \sum_\sigma\sum_{k\in\mathcal{K}_\sigma} n^\sigma_k \log\frac{n^\sigma_k}{N_\sigma \phi^{(k)}_f(t_\sigma, s_\sigma)}.
\end{equation}
The first sum runs over CG slices $\sigma$ and the second sum over all non-zero chunk sizes within that slice. $n^\sigma_k$ is the number of chunks of size $k$ in slice $\sigma$, and $N_\sigma$ the total number of chunks in slice $\sigma$. Finally $\phi^{(k)}_f(t_\sigma, s_\sigma)$ is the ML chunk distribution found in Eq.~\ref{eq:chunk}. Implementing an exact gof requires that we find the distribution of $G$ over all trees given the MLE of the model parameters and the constraint of a given tree depth with a given number of leaves, namely
\begin{equation}\label{eq:gof}
\text{Pr}(\text{exact}) = \sum_{\mathcal{T}'_K:G(\mathcal{T}'_K) \ge G(\mathcal{T}_K)} \text{Pr}(\mathcal{T}'_K)
\end{equation}
Doing the sum exactly would require exploring the space of all input data (trees) satisfying identical constraints. This would require varying all internal branch lengths, polytomy size/node degrees, and node positions in all possible ways while keeping the tree depth and size constant and is a combinatorially intractable task for even modestly sized trees. To approximate this intractable sum we will instead use a Metropolis-Hastings MCMC algorithm, which we describe next.

\subsection{CG proposal distribution}
Fortunately the CG representation of trees gives us a way to completely bypass the need to implement multiple types of proposal distributions (topological moves and metric moves) which would have to be followed by an expensive CGing step of the proposed tree. All those types of move can be subsumed into a single proposal distribution: the uniform random sampling of partitions. The intuition behind this proposal is as follows. If a node within a chunk is dragged across a slice interface towards the past, the chunk it was part of gets fragmented and the chunk further in the past into which the node moves increases in size. Moving a node across a slice interface towards the present has the opposite effect, namely it decreases the size of its chunk of origin and coagulates several chunks in its new slice. Topological moves restricted to a given slice simply reshuffle chunk order and has no effect on the chunk frequencies within a slice (for models with independent lineages). Arbitrary topological moves across slices are equivalent to a series of fragmentation and coagulation events and are therefore implicitly realized by the moves described above. Proposing a CG move will therefore consists of four steps:
\begin{itemize}
\item Choose a random chunk anywhere in the CG phylogeny except for the first slice which contains the leaves. Denote its size $k_s$.
\item Choose (without replacement) $k_s$ random chunks in the previous slice (immediately closer to the present). Their sizes are indicated by the partition (multiset) $\pi_s = \lbrace \lambda_j\rbrace_{j=1}^{k_s}$. This is a partition of $k = \sum_{\lambda\in\pi_s} \lambda$. Denote the multiplicity of parts in $\pi_s$ by $m_{\pi_s}(\lambda)$, $\lambda\in\mathbb{N}^+$.
\item Choose uniformly at random a partition $\pi_{t}$ of $k$ and let $k_t=\vert\pi_{t}\vert$ be the number of parts in $\pi_{t}$. Denote those parts $\lambda_n$ with $n$ running from 1 to $k_t$ and their multiplicities $m_{\pi_t}(\lambda)$, $\lambda\in\mathbb{N}^+$.
\item Replace $k_s$ by $k_t$, and the chunks with sizes in $\pi_s$ by the chunks with sizes in $\pi_t$. The number of elements $k_s=\vert\pi_s\vert$ and $k_t=\vert\pi_t\vert$ are not necessarily equal (there is no conservation rule at the interface across iterations), but the sums of their parts are equal because they are both partitions of the same number $k$ (conservation rule within an iteration).
\end{itemize}
Choosing randomly which $k_s$ chunks to repartition followed by the sampling a new partition with an arbitrary number $k_t$ of parts seamlessly combines both metric and topological moves, maintains fixed the depth of the tree and the number of leaves, and satisfies the lineage conservation constraint whereby the sum of chunk sizes in a slice is equal to the number of chunks in the adjacent slice closer to the present. We used the excellent and fast algorithm by Arrita and DeSalvo\cite{arratia_probabilistic_2011,desalvo_probabilistic_2014} to implement the uniform random generation of partition. We have the following algorithm:
\subsection{Metropolis-Hastings MCMC algorithm}\label{subseq:mcmc}
\begin{itemize}
\item Propose a move $\mathcal{T}'_K = \mathcal M[\mathcal{T}_K]$,
\item Accept move with probability $\alpha = \min\left[1, \frac{\text{Pr}[\mathcal{T}'_K]}{\text{Pr}[\mathcal{T}_K]}\frac{g\left(\mathcal{T}_K\vert \mathcal{T}'_K\right)}{g\left(\mathcal{T}'_K\vert \mathcal{T}_K\right)}\right]$,
\item Repeat last two steps and occasionally return a sample tree $\mathcal{T}_K$ and statistic $G(\mathcal{T}_K)$ according to a random sweep schedule.
\end{itemize}
We burn half the chain and return 100 samples. A sweep is considered complete when the number of proposed moves reaches about twice the number of chunks in the CG tree. We found this rule of thumb sufficient to eliminate autocorrelation between samples. The Metropolis ratio
\begin{equation}
\frac{\text{Pr}[\mathcal{T}'_K]}{\text{Pr}[\mathcal{T}_K]} = \frac{\phi_f^{(k_t)}(t_s, s_s)}{\phi_f^{(k_s)}(t_s, s_s)}.
\end{equation}
While the uniform random sampling of partitions is symmetric in the space of partitions it is not symmetric in the space of trees and we need to introduce the Hastings ratio
\begin{equation}
\frac{g\left(\mathcal{T}_K\vert \mathcal{T}'_K\right)}{g\left(\mathcal{T}'_K\vert \mathcal{T}_K\right)} = \frac{\vert\text{Comp}(\pi_s)\vert^{-1}}{\vert\text{Comp}(\pi_t)\vert^{-1}}= \frac{\vert\pi_t\vert!}{\prod_k m_{\pi_t}(k)!}\frac{\prod_k m_{\pi_s}(k)!}{\vert\pi_s\vert!}.
\end{equation}
To a given partition $\pi$ corresponds several underlying CG trees and those are enumerated by the set of compositions equivalent to a given partition. For a partition $\pi = \lbrace\lambda_n\rbrace_n$ with $\vert\pi\vert$ parts the number of equivalent compositions is given by the multinomial coefficient $\vert\text{Comp}(\pi)\vert = \vert\pi\vert!/\prod_n m_\pi(n)!$. For example, take the partition $\pi = (2, 1, 1, 1)$ of 5. This partition has $k = 4$ parts and multiplicities $m(1) = 3$, $m(2) = 1$, and $m(\lambda) = 0$ for all other $\lambda\not\in \lbrace 1, 2\rbrace$. Therefore $\vert\text{Comp}(\pi)\vert = 4!/(1!3!) = 4$. Indeed the set of equivalent compositions $\text{Comp}(\pi) = \lbrace(2, 1, 1, 1), (1, 2, 1, 1), (1, 1, 2, 1), (1, 1, 1, 2)\rbrace$.

The output of the MCMC is a set of IID trees $\mathcal{T}_K^1$, $\mathcal{T}_K^2$, $\mathcal{T}_K^3$, $\ldots$ distributed according the ML chunk size distribution of a model with all constraints satisfied, together with IID samples
$G(\mathcal{T}_K^1)$, $G(\mathcal{T}_K^2)$, $G(\mathcal{T}_K^3)$, $\ldots$ of the $G$-statistic. The Metropolis-Hastings algorithm effectively biases the MCMC walk in the space of CG trees toward trees that are in some sense "typical" for their size and depth given the model under consideration with parameters at their MLE. The $p$-value of the gof is approximated by the fraction of values of $G$ from the MCMC that are larger than the empirical value $G(\mathcal{T}_K)$ and quantifies how "typical" the empirical tree is. An asymptotically "perfect" tree, i.e. with an empirical chunk distribution that fits the theoretical chunk distribution exactly, would have a $G$-statistic of zero and $p$-value of 1. A "typical" tree should have a $p$-value fluctuating around 0.5. The $G$-statistic of typical trees will distribute according to a $\chi^2$ distribution. We did not find a way to calculate the number of degrees of freedom of this distribution \textit{a priori}, which would render moot the whole MCMC procedure presented above. Instead we can fit a $\chi^2$ distribution to the MCMC output knowing that the number of degree of freedoms $k = \langle G(\mathcal{T})\rangle_\text{MCMC}$. This gives a complementary way to find an asymptotic approximation to the $p$-value, e.g. in cases where the MCMC is computationally very expensive and one can only obtain a few samples of the $G$-statistic in reasonable time.

Fig~S4 shows the distribution of the gof p-values from a sub-sample of size 366 of all samples. Using the asymptotic p-values, we see that around 57\% of model fits pass the exact gof test at significance level 0.05. When we look at the gof in conjunction with the LRT between BDI and BDH, we see that BDH passes the gof and is favorably selected in 39\% of samples. BDI is not rejected and passes the gof in 18\% of samples. BDI is not rejected and fails the gof in 4\% of samples. Finally, and maybe more interestingly, BDH is favorably selected but does not pass the gof in 39\% of samples.

\bibliographystyle{vancouver}
\bibliography{refs}

\clearpage
\begin{figure}[h]
\includegraphics[scale=0.45]{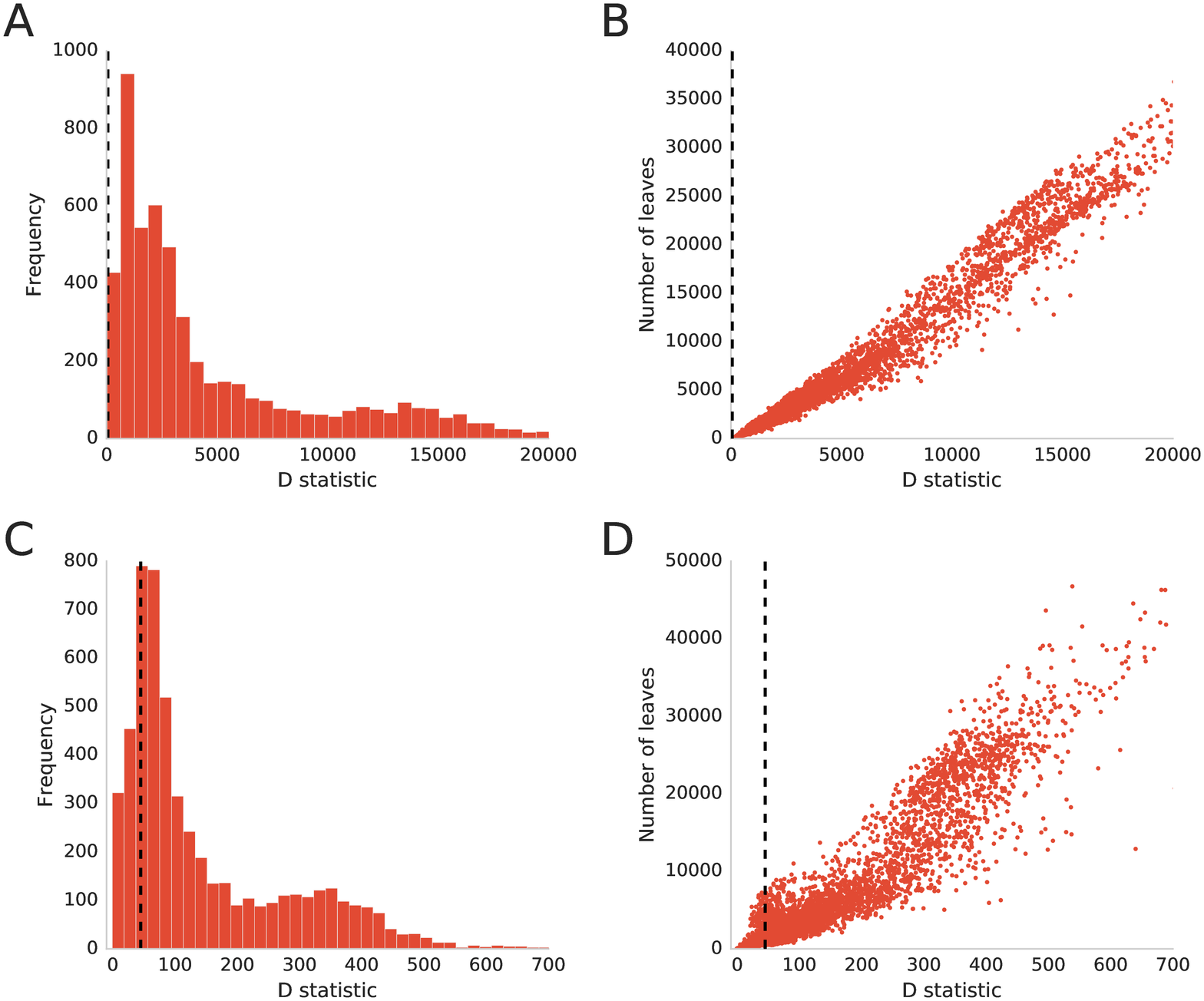}

\noindent{{\bf Fig S1. D statistic histograms for the LRT} ({\bf A}) Histogram for the D statistic when BD as the null hypothesis and BDI as the alternative. The black dashed line represent the D statistic at $5\sigma$ FWER with ddof=2. At this level BD is rejected in favour of BDI in 98\% of samples. ({bf B}) Relationship between the number of leaves in the sample tree and the D statistic for BD vs BDI. Each dot represents the LRT from one sample tree. ({\bf C}) and ({\bf D}) recapitulates the same results but for BDI vs BDH and therefore ddof=1. At this level BDI is rejected in favour of BDH in 80\% of samples. Notice in ({\bf B}) and ({\bf D}) how the rejection of BD vs BDI, and of BDI vs BDH strongly correlates with the size of the input tree---the inference picks up the signal from fast processes much more dramatically in large phylogenies.}
\end{figure}

\clearpage
\begin{figure}
\includegraphics[scale=0.53]{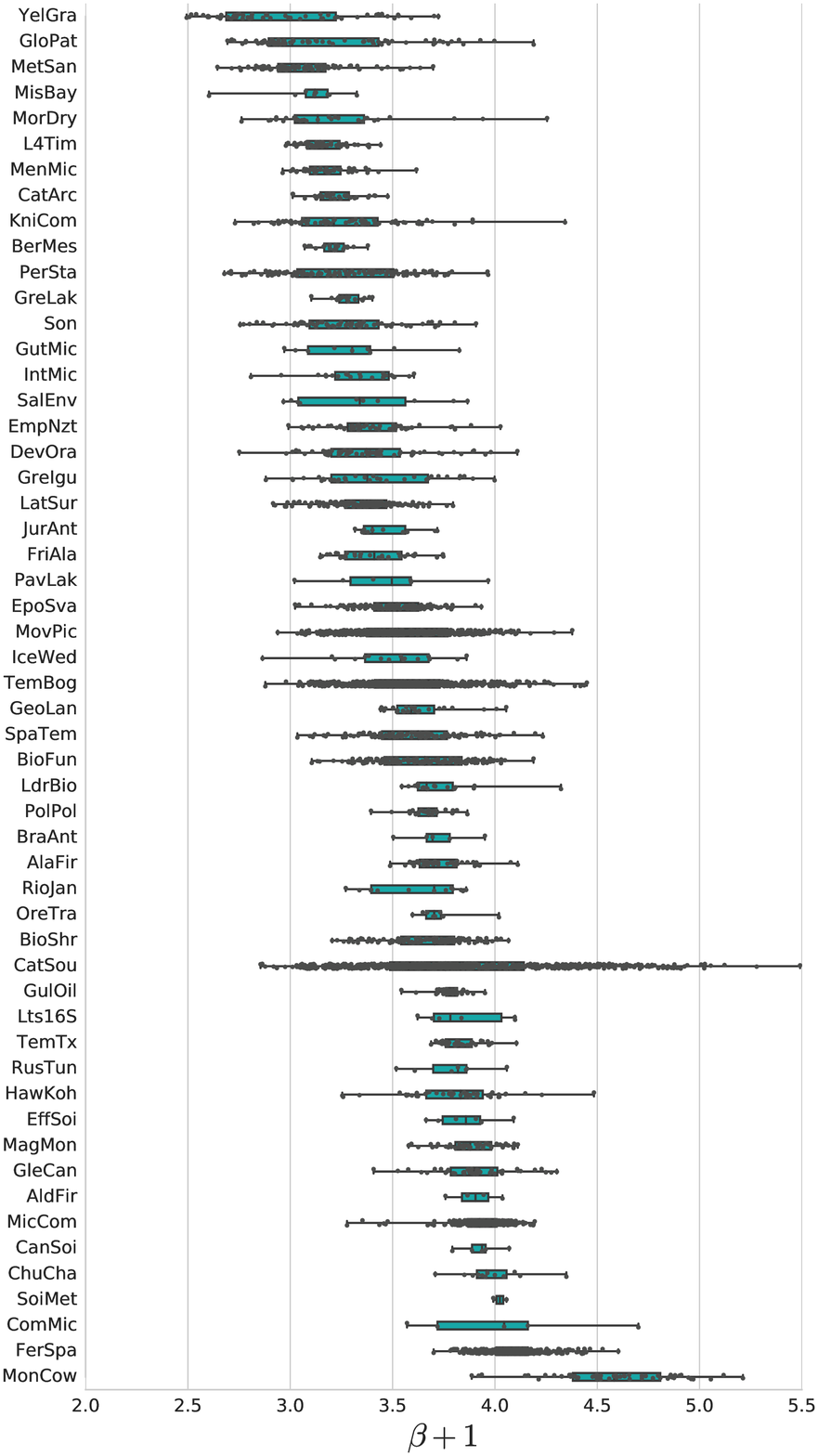}

\noindent{{\bf Fig S2. Tail exponent of the burst size distribution across studies.} See Table~S1 for the list of study abbreviations}
\end{figure}

\clearpage
\begin{figure}
\includegraphics[scale=0.52]{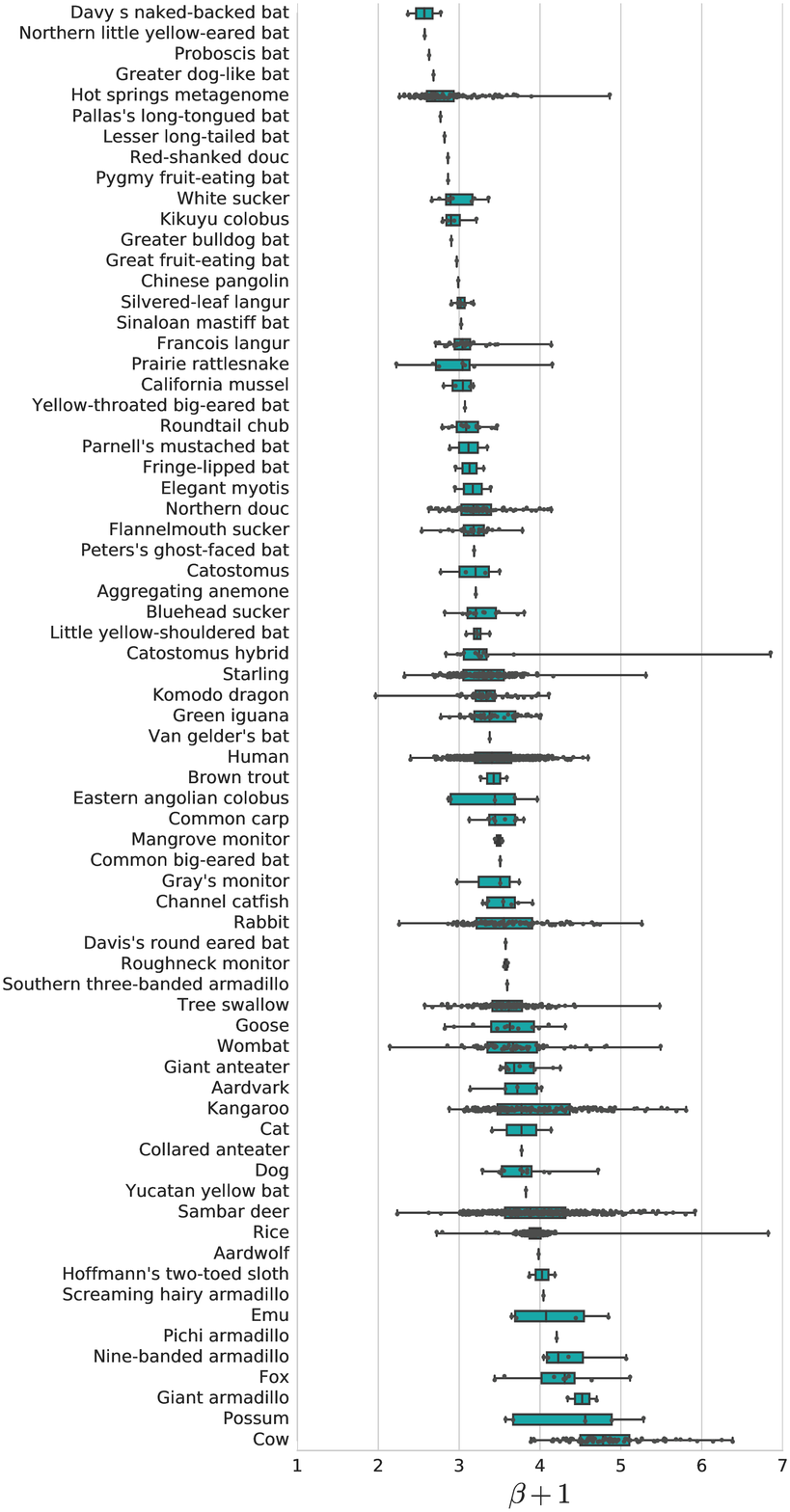}

\noindent{{\bf Fig S3. Tail exponent of the burst size distribution across host-associated microbiomes.}}
\end{figure}

\clearpage
\begin{figure}
\includegraphics[scale=0.54]{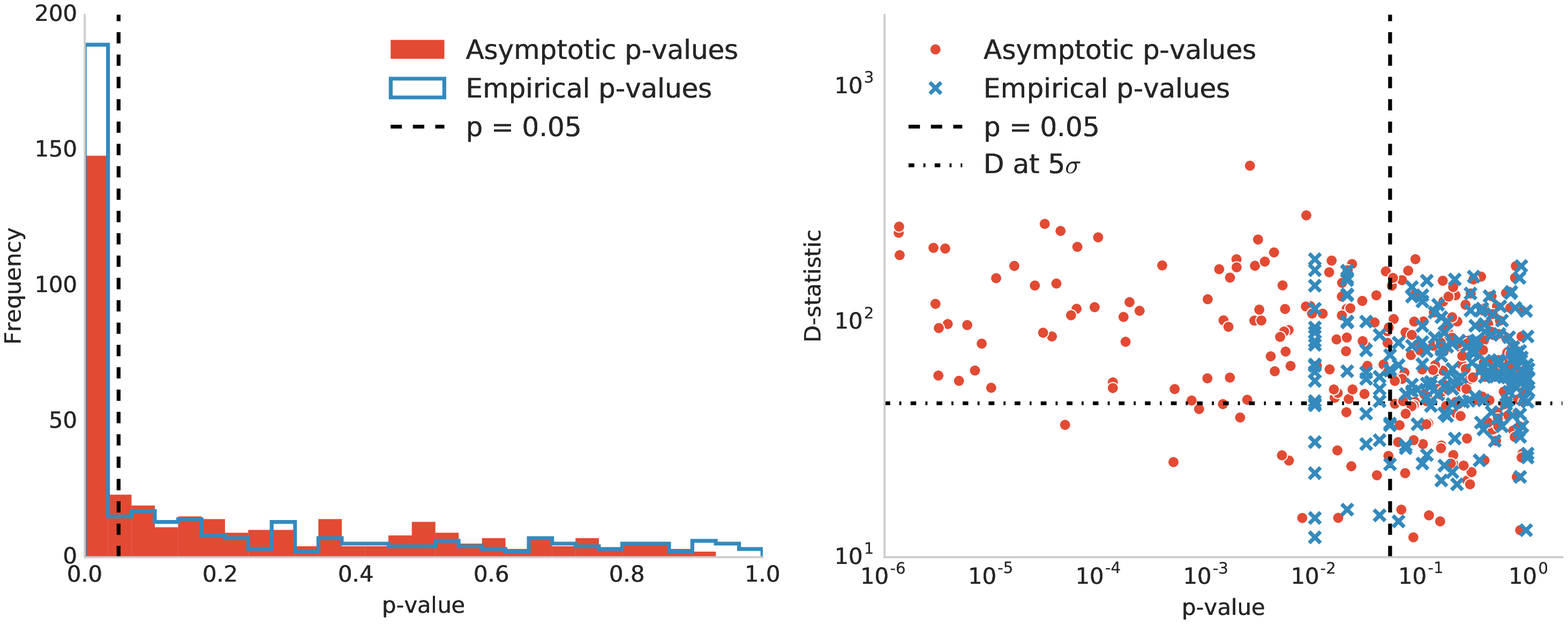}

\noindent{{\bf Fig S4. Distribution of goodness of fit p-values.} The histogram on the left shows the distribution of asymptotic and empirical gof p-values regardless of the best model. The fractions of asymptotic/empirical p-values falling to the right of the significance line are (0.62, 0.50). The scatter plot on the right show the same distribution vs the distribution of the D statistic for the LRT between BDI and BDH. The fraction of asymptotic/empirical (p, D) values falling within the quadrants are, starting from the top right corner and proceeding clockwise, (0.39, 0.33), (0.18, 0.14), (0.04, 0.08), and (0.39, 0.45).}
\end{figure}

\clearpage

\begin{table}
\resizebox*{!}{\dimexpr\textheight-2.55\baselineskip\relax}
{
\begin{tabular}{ r | p{7cm} | l | l}
	 Abbreviation & Short study description & Qiita ID & Accession/study link\\
	\hline
	{\bf AlaFir} & Alaskan Fire Chronosequence - Tanana Valley & \href{http://qiita.ucsd.edu/study/description/1030}{1030} & \url{http://www.ebi.ac.uk/ena/data/view/PRJEB14866}\\
	{\bf AldFir} & Alder/Fir & \href{http://qiita.ucsd.edu/study/description/1031}{1031} & \url{http://www.ebi.ac.uk/ena/data/view/PRJEB15055}\\
	{\bf BerMes} & Bergen Mesocosm & \href{http://qiita.ucsd.edu/study/description/1222}{1222} & \url{http://www.ebi.ac.uk/ena/data/view/PRJEB14793}\\
	{\bf BioFun} & Biodiversity And Functional Patterns Of Microbial Assemblages In Postglacial Pond Sediment Profiles & \href{http://qiita.ucsd.edu/study/description/1622}{1622} & \url{http://www.ebi.ac.uk/ena/data/view/PRJEB14823} \\
	{\bf BioShr} & Bioturbating Shrimp Alter The Structure And Diversity Of Bacterial Communities In Coastal Marine Sediments & \href{http://qiita.ucsd.edu/study/description/678}{678} & \url{http://www.ebi.ac.uk/ena/data/view/PRJEB15461} \\
	{\bf BraAnt} & Brazilian Antarctic Cleanup & \href{http://qiita.ucsd.edu/study/description/1033}{1033} & \url{http://www.ebi.ac.uk/ena/data/view/PRJEB14907}\\
	{\bf CanSoi} & Cannabis Soil Microbiome & \href{http://qiita.ucsd.edu/study/description/1001}{1001} & \url{http://www.ebi.ac.uk/ena/data/view/PRJEB15461}\\
	{\bf CatArc} & Catlin Arctic Survey 2010 & \href{http://qiita.ucsd.edu/study/description/723}{723} & \url{http://www.ebi.ac.uk/ena/data/view/PRJEB18098}\\
	{\bf CatSou} & Catchment Sources Of Microbes & \href{http://qiita.ucsd.edu/study/description/894}{894} & \url{http://www.ebi.ac.uk/ena/data/view/PRJEB14739}\\
	{\bf ComMic} & Comparison Of Microbial Flora In Ant-Eating Mammals & \href{http://qiita.ucsd.edu/study/description/1056}{1056} & \url{http://www.ebi.ac.uk/ena/data/view/ERX301013}\\
	{\bf DevOra} & Development Of The Oral Microbiota In Captive Komodo Dragons (Varanus Komodoensis) & \href{http://qiita.ucsd.edu/study/description/1747}{1747} & \url{http://www.ebi.ac.uk/ena/data/view/PRJEB14602} \\
	{\bf EffSoi} & Effect Of Soil Ph On Soil Metagenome & \href{http://qiita.ucsd.edu/study/description/805}{805} & \url{http://www.ebi.ac.uk/ena/data/view/PRJEB18597} \\
	{\bf EmpNzt} & EMP Nztabs & \href{http://qiita.ucsd.edu/study/description/1035}{1035} & \url{http://www.ebi.ac.uk/ena/data/view/PRJEB3228}\\
	{\bf EpoSva} & EPOCA Svalbard & --- & \url{https://doi.pangaea.de/10.1594/PANGAEA.769833} \\
	{\bf FerSpa} & Fermilab Spatial Study & --- & ---\\
	{\bf FriAla} & Friedman Alaska Peat Soils & \href{http://qiita.ucsd.edu/study/description/1692}{1692} & \url{http://www.ebi.ac.uk/ena/data/view/PRJEB15218}\\
	{\bf GeoLan} & Geochemical Landscapes & \href{http://qiita.ucsd.edu/study/description/1036}{1036} & \url{http://www.ebi.ac.uk/ena/data/view/PRJEB14909}\\
	{\bf GleCan} & Glen Canyon Soils & \href{http://qiita.ucsd.edu/study/description/1526}{1526} & \url{http://www.ebi.ac.uk/ena/data/view/PRJEB15163}\\
	{\bf GloPat} & Global Patterns Of 16S Rrna Diversity At A Depth Of Millions Of Sequences Per Sample & \href{http://qiita.ucsd.edu/study/description/721}{721} & \url{http://www.ebi.ac.uk/ena/data/view/PRJEB18917}\\
	{\bf GreIgu} & Green Iguana Hindgut Microbiome & \href{http://qiita.ucsd.edu/study/description/963}{963} & \url{http://www.ebi.ac.uk/ena/data/view/PRJEB15058}\\
	{\bf GreLak} & Great Lake Microbiome & \href{http://qiita.ucsd.edu/study/description/1041}{1041} & \url{http://www.ebi.ac.uk/ena/data/view/PRJEB14819}\\
	{\bf GulOil} & Gulf Oil Spill Sediment & \href{http://qiita.ucsd.edu/study/description/1197}{1197} & \url{http://www.ebi.ac.uk/ena/data/view/PRJEB14900}\\
	{\bf GutMic} & Gut Microbiota Of Phyllostomid Bats That Span A Breadth Of Diets & \href{http://qiita.ucsd.edu/study/description/1734}{1734} & \url{http://www.ebi.ac.uk/ena/data/view/PRJEB14489}\\
	{\bf HawKoh} & Hawaii Kohala Volcanic Soils & \href{http://qiita.ucsd.edu/study/description/1579}{1579} & \url{http://www.ebi.ac.uk/ena/data/view/PRJEB15174}\\
	{\bf IceWed} & Ice Wedge Polygon & \href{http://qiita.ucsd.edu/study/description/1578}{1578} & \url{http://www.ebi.ac.uk/ena/data/view/PRJEB9043}\\
	{\bf IntMic} & Intertidal Microbes & \href{http://qiita.ucsd.edu/study/description/662}{662} & \url{http://www.ebi.ac.uk/ena/data/view/PRJEB18565}\\
	{\bf JurAnt} & Jurelivicius Antarctic Cleanup & \href{http://qiita.ucsd.edu/study/description/776}{776} & \url{http://www.ebi.ac.uk/ena/data/view/PRJEB15611}\\
	{\bf KniCom} & Knight Comp Biogeography &\href{http://qiita.ucsd.edu/study/description/1748}{1748} & \url{http://www.ebi.ac.uk/ena/data/view/ERP022166}\\
	{\bf L4Tim} & L4 Time Series 2009-2010 & \href{http://qiita.ucsd.edu/study/description/1240}{1240} & \url{http://www.ebi.ac.uk/ena/data/view/PRJEB14864}\\
	{\bf LatSur} & Latitudinal Surveys Of Algal-Associated Microorganisms & \href{https://qiita.ucsd.edu/study/description/933}{933} & \url{http://www.ebi.ac.uk/ena/data/view/ERP021699}\\
	{\bf LdrBio} & Ldrd Biological Carbon Sequestration & \href{http://qiita.ucsd.edu/study/description/1043}{1043} & \url{http://www.ebi.ac.uk/ena/data/view/PRJEB14912} \\
	{\bf Lts16S} & Ltsp & \href{http://qiita.ucsd.edu/study/description/1037}{1037} & \url{http://www.ebi.ac.uk/ena/data/view/PRJEB14908}\\
	{\bf MagMon} & Magnificent Mongolian Microbes & \href{http://qiita.ucsd.edu/study/description/864}{864} & \url{http://www.ebi.ac.uk/ena/data/view/PRJEB15460}\\
	{\bf MenMic} & Mendota Microbial Observatory & \href{http://qiita.ucsd.edu/study/description/1242}{1242} & \url{http://www.ebi.ac.uk/ena/data/view/PRJEB14911}\\
	{\bf MetSan} & Metcalf Sandiego Zoo Folivorus Primate & \href{http://qiita.ucsd.edu/study/description/2182}{2182}	 & \url{http://www.ebi.ac.uk/ena/data/view/PRJEB14631}\\
	{\bf MicCom} & Microbial Community Of The Bulk Soil And Rhizosphere Of Rice Plants Over Its Lifecycle & \href{http://qiita.ucsd.edu/study/description/1642}{1642} & \url{http://www.ebi.ac.uk/ena/data/view/PRJEB15194}\\
	{\bf MisBay} & Mission Bay Sediment Viromes & \href{http://qiita.ucsd.edu/study/description/1673}{1673} & \url{http://www.ebi.ac.uk/ena/data/view/PRJEB15214}\\
	{\bf MonCow} & Monensin Cow Hindgut Study Cornell & \href{http://qiita.ucsd.edu/study/description/1621}{1621} & \url{http://www.ebi.ac.uk/ena/data/view/PRJEB14795}\\
	{\bf MorDry} & Morgankiss Dry Valley Lake Communities Protist Diversity In A Permanently Ice-Covered Antarctic Lake During The Polar Night Transition & \href{http://qiita.ucsd.edu/study/description/638}{638} & \url{http://www.ebi.ac.uk/ena/data/view/PRJEB18567}\\
	{\bf MovPic} & Moving Pictures Of The Human Microbiome & \href{http://qiita.ucsd.edu/study/description/550}{550} & \url{http://www.ebi.ac.uk/ena/data/view/PRJEB19825}\\
	{\bf OreTra} & Oregon Transect & \href{http://qiita.ucsd.edu/study/description/1038}{1038} & \url{http://www.ebi.ac.uk/ena/data/view/PRJEB14862} \\
	{\bf PavLak} & Pavilion Lake Research Project & \href{http://qiita.ucsd.edu/study/description/809}{809} & \url{http://www.ebi.ac.uk/ena/data/view/PRJEB18097}\\
	{\bf PerSta} & Peralta Starlings & \href{http://qiita.ucsd.edu/study/description/1694}{1694} & \url{http://www.ebi.ac.uk/ena/data/view/PRJEB14798}\\
	{\bf PolPol} & Polluted Polar Coastal Sediments & \href{http://qiita.ucsd.edu/study/description/1198}{1198} & \url{http://www.ebi.ac.uk/ena/data/view/PRJEB14880}\\
	{\bf RioJan} & Rio De Janeiro Coastline & \href{http://qiita.ucsd.edu/study/description/1039}{1039} & \url{http://www.ebi.ac.uk/ena/data/view/PRJEB15044} \\
	{\bf RusTun} & Russian Tundra Samples Cryocarb & \href{http://qiita.ucsd.edu/study/description/1034}{1034} & \url{http://www.ebi.ac.uk/ena/data/view/PRJEB15045}\\
	{\bf SalEnv} & Saline Environments That May Harbor Novel Lignocellulolytic Activities Tolerant Of Ionic Liquids & \href{http://qiita.ucsd.edu/study/description/1580}{1580} & \url{http://www.ebi.ac.uk/ena/data/view/PRJEB15178} \\
	{\bf SoiMet} & Soil Metagenome & \href{http://qiita.ucsd.edu/study/description/808}{808} & \url{http://www.ebi.ac.uk/ena/data/view/PRJEB18643} \\
	{\bf Son} & Microbiota of freshwater fish slime and gut & \href{http://qiita.ucsd.edu/study/description/940}{940} & \url{http://www.ebi.ac.uk/ena/data/view/PRJEB14822} \\
	{\bf SpaTem} & Spatial And Temporal Variation In Nest And Egg Bacteria Of Wild Birds & \href{http://qiita.ucsd.edu/study/description/1098}{1098} & \url{http://www.ebi.ac.uk/ena/data/view/PRJEB15144}\\
	{\bf TemBog} & Temperate Bog Lakes & \href{http://qiita.ucsd.edu/study/description/1288}{1288} & \url{http://www.ebi.ac.uk/ena/data/view/PRJEB15148}\\
	{\bf TemTx} & Temple Tx Native Exotic Ppt  & \href{http://qiita.ucsd.edu/study/description/1289}{1289} & \url{http://www.ebi.ac.uk/ena/data/view/PRJEB15146}\\
	{\bf YelGra} & Yellowstone Gradients & \href{http://qiita.ucsd.edu/study/description/925}{925} & \url{http://www.ebi.ac.uk/ena/data/view/PRJEB20056}\\
	\hline
\end{tabular}
}

\noindent{{\bf Table S1. List of studies included in the analysis and their abbreviations, Qiita ID, and accession link.}}
\end{table}

\end{document}